\documentclass[twocolumn]{aastex63}
\usepackage{amsmath}
\usepackage{graphicx}
\usepackage{xcolor}
\usepackage{hyperref}
\usepackage{outlines}
\usepackage{tablefootnote}
\usepackage{booktabs}

\usepackage{mathtools}
\usepackage[caption=false]{subfig}

\newcommand{\beq}{\begin{equation}}
\newcommand{\eeq}{\end{equation}}
\newcommand{\bea}{\begin{eqnarray}}
\newcommand{\eea}{\end{eqnarray}}
\newcommand{\mrm}{\mathrm}

\newcommand{\black}{\color{black}}

\usepackage[nolist,nohyperlinks]{acronym}
\newacro{ISM}{interstellar medium}
\newacro{VLBI}{very long baseline interferometry}
\defcitealias{Rickett2014}{R14}

\usepackage{xspace}
\def\HII{\ion{H}{2}\xspace}
\def\HI{\ion{H}{1}\xspace}

\shorttitle{Scintillation and ISM Filaments}
\shortauthors{Stock \& van Kerkwijk}

\begin{document}

\title{ Associations Between Scattering Screens and Interstellar Medium Filaments}

\correspondingauthor{Ashley M. Stock}
\email{ashley.stock@mail.utoronto.ca}

\author[0000-0001-5351-824X]{Ashley M. Stock}
\affiliation{David A. Dunlap Department of Astronomy \& Astrophysics, University of Toronto, 50 Saint George St., Toronto, ON M5S 3H4, Canada}
\affiliation{Canadian Institute for Theoretical Astrophysics, University of Toronto, 60 Saint George Street, Toronto, ON M5S 3H8, Canada}

\author[0000-0002-5830-8505]{Marten H. van Kerkwijk}
\affiliation{David A. Dunlap Department of Astronomy \& Astrophysics, University of Toronto, 50 Saint George St., Toronto, ON M5S 3H4, Canada}

\begin{abstract}
 Pulsar scintillation can be used to measure small scale structure in the Galaxy, but little is known about the specific interstellar medium features that cause scintillation.
  We searched for interstellar medium counterparts to all scintillation screens for which absolute distances and scattering orientations have been measured – a sample of 12 pulsars and 22 screens.
  For one pulsar, PSR J0737-3039A, we re-analyze its scintillation screen and find evidence for a highly anisotropic screen.
  Comparisons with tracers of ionized media did not find any new associations.
  Instead, for seven of the pulsars analyzed, aligned \HI filaments are seen for at least one of their screens, for a total of 12 out of 22 screens.
  This result seems unlikely to be due to chance: comparing with random trials, we estimate a likelihood of finding 12 or more screens with aligned emission by chance of only 0.004\%.
  Estimating the significance of the amount of aligned emission (in standard deviations over the mean), the probability of finding as much observed aligned emission by chance is larger, at 1.7\%, but still indicative of a real correlation.  Since \HI filaments are preferentially associated with cold neutral gas, and thus unlikely to cause scintillation, this may indicate both the filaments and the screens are aligned preferentially by the same mechanisms such as magnetic fields or shocks.
\end{abstract}


\section{Introduction}\label{Sec:Intro}

The \ac{ISM} contains a rich assortment of structures such as clouds, filaments, shells, and bubbles.
The size distribution of the structures has long been known to follow a non-uniform power law
\citep[e.g.][]{Armstrong1981,Armstrong1995,Chepurnov2010}.
At large scales, from $10^{16}$ to $10^{18}{\rm\;m}$, structures are observed directly, and the power law has been observed across multiple phases of the ISM, with the emission of cold, neutral hydrogen \citep[e.g.][]{Lazarian2000} as well as warm, ionized hydrogen \citep[e.g.][]{Chepurnov2010}.
Below $10^{14}{\rm\,m}$ the distributions are not observed directly, except in the very local ISM environment by probes such as Voyager 1 and 2 \citep[e.g.][]{Lee2019}, but are rather inferred from the scattering of compact radio sources, such as pulsars and the cores of active galactic nuclei.
From those indirect measurements, the distribution has been suggested to extend all the way down to $10^6{\rm\;m}$ \citep{Armstrong1981}.

Compact radio sources are powerful probes of the ISM because they emit spatially coherent light.
As the light propagates through the inhomogeneous ISM, it is scattered along many paths that interfere with each other, causing the intensity and phase of the light to vary with frequency and time as the source, observer, and scattering material move relative to each other.
These variations, collectively known as scintillation, thus encode information about material along the line-of-sight to the source.

Scintillation is typically observed through the dynamic spectrum, which shows mean intensity as a function of frequency and time.
Bright regions, called ``scintles'', have typical frequency widths of several kHz to MHz and durations on the order of minutes for pulsars and weeks for active galactic nuclei (AGN), when observed at GHz frequencies.
The characteristic widths are quantified as the ``decorrelation bandwidth'' and ``scintillation timescale'', respectively \citep[e.g.,][]{Cordes1985}.
Occasionally, ``extreme scattering events'' are seen \citep[e.g.][]{Fiedler1987},
large changes in intensity that last much longer than the scintillation timescale and cover a much wider frequency range than the decorrelation bandwidth (indeed, they often are  accompanied by changes in the scintillation timescale and decorrelation bandwidth).

Although scintillation gives information about the electron density and size distribution of the scattering medium, the nature of structures probed by scintillation remains poorly understood.
Scintillation must occur in partially ionized gas, requiring electron densities likely higher than those in the molecular (too cold) and hot ionized (too tenuous) phases of the \ac{ISM}, i.e., most likely in the warm neutral or warm ionized phases.

Clear parabolic structures in ``secondary spectra'', the power spectra of dynamic spectra, and, more directly, interferometric observations suggest the scattering is dominated by small numbers of localized, elongated structures  \citep[e.g.,][]{Rickett2009, Brisken2010}.
These structures are typically referred to as ``screens'', which may take the form of sheets or filaments.
Sheet-like structures are particularly attractive, since if they are aligned nearly parallel to the line-of-sight, they appear filamentary and also solve the over-pressure problem, i.e. that extreme scattering events require bending angles of order 10s of mas, which implies column density gradients of order $1000{\rm\;cm^{-2}\,cm^{-1}}$ \citep{Clegg1998}, and thus, for spherically symmetric structures, electron densities well higher than any expected in the \ac{ISM}.
In contrast, for a sheet sufficiently aligned with the line-of-sight, typical \ac{ISM} electron densities of $\sim\!1{\rm\;cm^{-3}}$ would suffice to cause the same bending angles \citep{Pen2014}.

%{\red \citep{Hill2005} -- bending angles of up to 20 mas}

Under the assumption of thin, elongated screens, it is possible to model the distances, orientations of elongation, and velocities of screens.
These properties can then be compared with properties of the ISM to potentially establish connections.
Surveys of ISM tracers lack the resolution needed to observe directly structures on the scales that cause scintillation, but if the scattering screen is part of a larger scale structure, associations might still be found.

Early attempts to make associations used only sky positions and scattering strengths of the screens.
For instance, extreme scattering events of quasars seem to be preferentially located near Galactic synchrotron filaments (loops) and near the edges of shell-like \HI structures within those filaments \citep{Fiedler1994}.
Scattering also appears enhanced for pulsars in the direction of Galactic Loop~I, possibly due to interaction between Loop~I and the Local Bubble \citep{Bhat2002}.

Using both sky positions and velocities of screens, \citet{Walker2017} found evidence that some screens were associated with hot stars.
Specifically, they found that two variable sources, PKS~1322$-$110 and J1819$+$3845, had sight lines that passed within 2 pc of hot stars and for which the inferred velocities of the screens matched those of the stars.
\citet{Walker2017} suggested that small, over-dense neutral clouds with tails ionized by hot stars might be responsible for the scintillation.

For a few pulsars, more direct associations between screens and specific ISM features have been made.
The Crab pulsar, Vela pulsar, and PSR~J0538$+$2817 are clearly scattered by their associated supernova remnants \citep{Cordes2004,Sushch2011,Yao2021}.
One of the scattering screens towards PSR~J1643$-$1224 is likely associated with an \HII region, Sh 2$-$27 \citep{Mall2022,Ding2023}, while three of the scattering screens of PSR\;B1133$+$16 \citep{McKee2022} and one of the scattering screens of B0656+14 \citep{Yao2022} appear associated with the edge of the Local Bubble.

Nevertheless, the majority of scintillation screens do not have a known association.  Additionally, using telescopes with higher sensitivity and resolution, there is increasing evidence that many scintillation screens can be observed along a given line-of-sight \citep[e.g.][]{Ocker2024,Reardon2024} suggesting they are more common than previously thought.  This suggests that scintillation screens should be associated with ISM structures, or a combination of structures, that are ubiquitous.

A typically unused parameter for identifying associations between screens and ISM features is the orientation of the scintillation screens.
Measuring the orientation of screens generally requires interferometric techniques \citep[e.g.][]{Brisken2010}, or observations over several months or years to fit for the annual and/or orbital variation in scattering properties \citep[e.g.][]{Rickett2014,Reardon2019,McKee2022}.
As a result, this has been done for very few pulsars so far.
However, this measurement is needed to constrain the inclination of binary pulsars and could potentially be used to reduce errors in measurements of parallax by correcting for the positional shifts induced by the scattering.

In this paper, we use distances and orientations to investigate connections between scintillation screens and structures in the ISM, by correlating screen orientations with morphological features in tracers of different ISM environments.
In Section \ref{section:Methods}, we outline the screens and ISM tracers used for comparison, and describe how we test for alignment.
We discuss the screens for each pulsar in detail in Section \ref{section:psr-idv}, and consider the results from all screens in Section \ref{section:psr-agg}.

\section{Data and Analysis}\label{section:Methods}

\subsection{Scattering Screens}

We collected information on scattering screens from the literature, limiting ourselves to those with measurements of both distance and orientation of anisotropy.
For all sources, we carefully verified how orientation is defined, and for PSR J0737$-$3039A we made new fits to existing scintillation measurements (see Appendix for details).
The results are summarized in Table~\ref{table:psr-lit-review}.
Further information can be found online in what we hope will become a live database\footnote{https://github.com/scintillometry/screen-database}, to which further screen distances and orientations will be added, such as those expected from the ongoing scintillation survey with the Thousand Pulsar Array \citep{Main2023}.

\begin{figure}
  \centering
  \includegraphics[width=\linewidth]{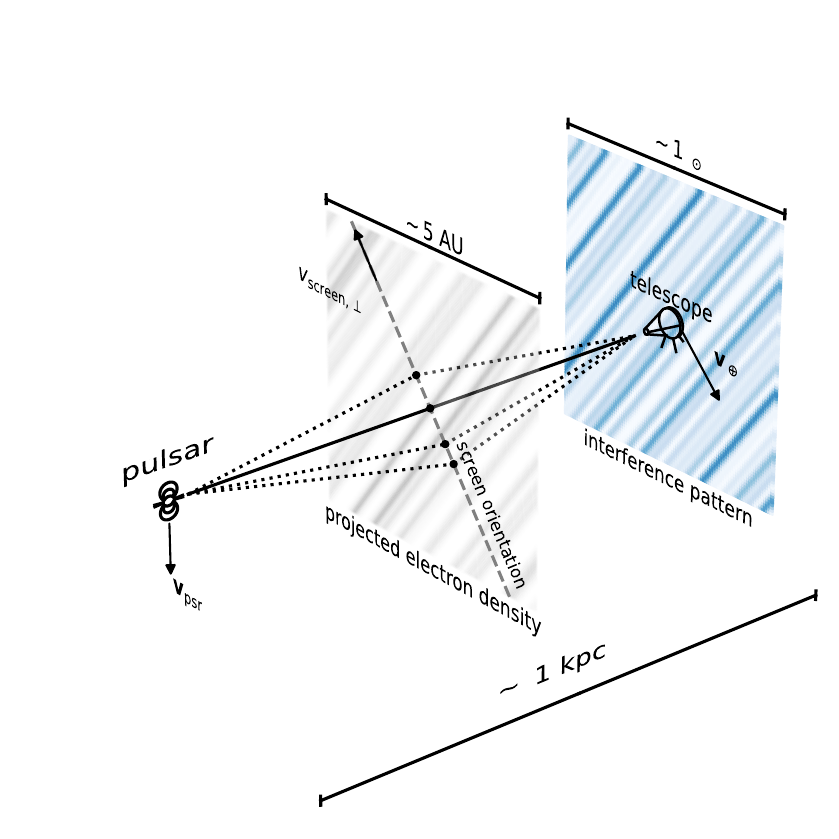}
  \caption{Schematic illustrating a possible system with a pulsar observed by a telescope through a screen, where the scintillation screen is aligned with an ISM filamentary or sheet-like density structure.
      Images of the pulsar form on the screen at regions with large spatial derivatives of column density---here depicted as overdense, elongated regions when seen in projection, similar to what is expected in the corrugated sheet model of \citet{Pen2014}, where the elongated structures correspond to waves on a sheet seen at grazing incidence.
      The screen orientation is defined by the average direction on the plane of the sky between images formed and the line of sight, which is perpendicular to the orientation of elongation of density structures.
      Only scattering points within a few mas of the line of sight can bend light sufficiently to reach the observer, creating the interference patterns observed as scintillation.
      The paths for only four of the scattering points are shown.
      The relative motion of the source, screen, and observer affect the temporal variation of the observed interference pattern but for the one-dimensional screen shown only motion in the direction of the screen orientation contributes.}
    \label{fig:filament-diagram}
\end{figure}

\subsubsection{Screen Orientations}

The screen orientations in Table \ref{table:psr-lit-review} are given East of North, and are measured along the long axis of the distribution of scattered images on the sky.
This filament orientation equals that inferred from direct imaging and tracing annual arc curvature measurements; it corresponds to the direction of steepest change in density and phase.
The orientation is perpendicular to what is inferred from scintillation timescales, which measure the direction associated with the longest scintillation timescale and thus that of least change in density and phase. This is illustrated in Figure \ref{fig:filament-diagram}.

For an elongated set of scattering points, there is necessarily a $180^\circ$ ambiguity in the orientation angle.
We have chosen a convention in which the angle is between 0 and $180^\circ$ so that the angle is towards the Eastern side of the long axis of the scattering points, adjusting the sign of the screen velocity appropriately if the original measurement from the literature followed a different convention (i.e. we ensured that the screen velocity is positive if it has a positive component in right ascension).

\subsubsection{Screen Kinematics}

For most of the pulsars, information on the velocity of the screen is available.
We only use the component of the velocity parallel to the structure's scattering points (see Table~\ref{table:psr-lit-review}), since for elongated distributions of scattering points, there is little change in scintillation as one moves perpendicular to them, and hence any velocity in that direction is highly uncertain.
Indeed, in some models for scintillation, such as those that require highly aligned and elongated sheets \citep{Pen2014}, apparent variations perpendicular to the long axis of the scattering points require the presence of at least one other sheet with a different orientation.

For comparison with the measured velocities, we calculate the expected velocity for a structure following bulk Galactic rotation relative to the Local Standard of Rest, listing in Table~\ref{table:psr-lit-review} both the component parallel to the screen scattering points and the radial velocity -- the latter for comparison with frequency-resolved \ac{ISM} tracers.
We calculate the expected value for the proper motion and kinematic Local Standard of Rest velocity for a screen as:
\begin{align}
    v_{\ell} =& d_s [B+A \cos(2\ell) - C \sin(2\ell)]\cos(b) \nonumber \\
    &+ U_0\sin(\ell) - V_0 \cos(\ell),\\
    v_b =& - d_s [K + A \sin(2\ell) + C \cos(2\ell)]\sin(b)\cos(b) \nonumber \\
    & + [U_0\cos(\ell) + V_0 \sin(\ell)]\sin(b) - W_0 \cos(b),\\
    v_{\mathrm{LSRK}} =& d_s [K + A \sin(2 \ell) + C \cos(2\ell)] \cos(b),
\end{align}
where $d_s$ is the distance to the screen from the Earth, and ($\ell, b$) are its Galactic longitude and latitude.
We used the Oort constants from \citet{Bovy2017},
\begin{equation}
  \left(\begin{matrix}A\\B\\C\\K\end{matrix}\right)
  = \left(\begin{matrix}
      15.3 \pm 0.4 \\
      -11.9 \pm 0.4\\
      -3.2 \pm 0.4\\
      -3.3 \pm 0.6\end{matrix}\right){\rm km\,s^{-1}\,kpc^{-1}},
\end{equation}
and the velocity of the Sun relative to the Local Standard of Rest from \citet{Schonrich2010},
\begin{equation}
  \left(\begin{matrix}U_0\\V_0\\W_0\end{matrix}\right)
  = \left(\begin{matrix}
      11.1^{+0.69}_{-0.75}\\
      12.24^{+0.47}_{-0.47}\\
      7.25^{+0.37}_{-0.36}\end{matrix}\right) {\rm km\,s^{-1}}.
\end{equation}

\begin{deluxetable*}{l c c c c c c c}[t]
\tabletypesize{\footnotesize}
\tablecaption{\label{table:psr-lit-review}  Summary of associations between literature pulsar screens and \HI filaments}
\tablehead{
&  & Screen & Screen  & Galactic & Screen &  Predicted \HI &\\
Pulsar & & Orientation & Velocity & Rotation & Distance & Radial Velocity & \\
Name & Screen & ($^\circ$) & (km/s) & (km/s) & (pc)  & (km/s) & Reference}
\startdata
J0337$+$1715  & a & $66.1_{-3.7}^{+3.7}$     & $4.25_{-0.42}^{+0.44}$   & $-0.369 \pm 0.011$ &{$244_{-3.1}^{+5.1}$} & $-2.44 \pm 0.05$& \citet{Gusinskaia_inprep} \\
J0437$-$4715  & a & $134.6 \pm 0.3$         & $-31.9 \pm 0.3$        & $6.37 \pm 0.03$    &{ $89.8 \pm 0.4$}    & $0.518 \pm 0.002$& \citet{Reardon2020}\\
              & b &   $144 \pm 6$           & $-50 \pm 6$            & $-6.08 \pm 0.19$   &{ $124  \pm 3$}      & $0.715 \pm 0.017$&\\
J0613$-$0200  &2014+&  $16 \pm 2$           & $-1.2 \pm 2.5$         & $-5.9 \pm 0.8$     &{ $300 \pm 50$}      & $2.5 \pm 0.4$& \citet{Main2020} \\
              &2013&  $144 \pm 9$           & $-12.8 \pm 2.8$        & $3.3 \pm 0.7$      &{ $330 \pm 100$}     & $2.8 \pm 0.8$&\\
J0636$+$5128  & a & $96.98^{+3.16}_{-3.63}$  & $-2.01^{+0.97}_{-1.15}$  & $-20 \pm 5$        &{ $262^{+98}_{-38}$}  & $-3.5 \pm 1.3$& \citet{Liu2023}\\
J0737$-$3039A & a &  $88.6 \pm 4.3$         & $-25.6 \pm 1.5$        & $3.80 \pm 0.04$    &{ $260 \pm 51$}      & $2.7 \pm 0.5$& This Work \\
B0834$+$06    & a &   $155 \pm 1$           &  $-23 \pm 1$           & $-12.94 \pm 0.13$  &{\black  $389 \pm 5$}& $3.89 \pm 0.05$&\citet{Zhu2023}\\ %\citet{Zhu_inprep}
              & b &   $136 \pm 1$           & $-3.0 \pm 3$           & $-12.1 \pm 0.2$    &{\black $415 \pm 11$}& $4.15 \pm 0.11$&\\
B1133$+$16    & b & $154^{+55}_{-37}$        & $101^{+98}_{-169}$      & $4.4 \pm 1.2$      &{  $186^{+82}_{-87}$} & $0.7 \pm 0.4$& \citet{McKee2022}\\
              & c & $120^{+24}_{-23}$        & $78^{+88}_{-47}$        & $0.13 \pm 0.07$    &{  $134^{+27}_{-31}$} & $0.54 \pm 0.12$& \\
              & d & $128^{+51}_{-39}$        & $-399^{+230}_{-135}$    & $8.4 \pm 1.9$      &{  $197^{+67}_{-59}$} & $0.8 \pm 0.3$& \\
              & e & $135^{+50}_{-38}$        & $-74^{+173}_{-142}$     & $5.5 \pm 1.3$      &{  $186^{+69}_{-73}$} & $0.7 \pm 0.3$& \\
              & f & $11.3^{+2.4}_{-2.3}$     & $-5.28^{+0.37}_{-0.42}$ & $-6.1 \pm 0.6$      &{$5.47^{+0.54}_{-0.59}$}&$0.022 \pm 0.002$&\\
J1141$-$6545  & a & $28.0 \pm 0.4$          & \nodata                & $91.4 \pm 0.8$     &{\black$7200^{+300}_{-220}$}& $-96 \pm 4$& \citet{Reardon2019}\\
B1508$+$55\tablenotemark{a}
       & a (weak) & $39.8 \pm 2.0 $         & $7.3 \pm 1.4$          & $3.86 \pm 0.06$    &{ $127.1 \pm 1.5$}   & $-0.0630 \pm 0.0007$&\citet{Sprenger2022}\\
       & a (strong) & $52.4 \pm 0.4$\\
              & b & $129.8 \pm 2.0$         & $2 \pm 55$             & $10.5$             &{\black $1940 \pm 123$}& $-0.96 \pm 0.06$&\\
J1603$-$7202  & a & $140 \pm 13$            & $-5^{+15}_{-18}$        & $-41 \pm 4$        &{\black $2500 \pm 700$}& $-45 \pm 13$ &\citet{Walker2022}\\
J1643$-$1224  & a & $150 \pm 4$             & $-10    \pm 1$         & $9.1 \pm 0.9$      &{ $129 \pm 15$}\tablenotemark{b}& $-0.41 \pm 0.05$ & \citet{Mall2022}\\
              & b & $31 \pm 9$              & $-6 \pm 3$             & $-3.20 \pm 0.11$   &{\black $340 \pm 9$} & $-1.09 \pm 0.03$& \\
J1909$-$3744  & a & $85 \pm 6 $             & $14 \pm 10$            & $14.3 \pm 1.4$     &{\black  $590 \pm 50$}& $-3.7 \pm 0.3$&\citet{Askew2023}\\
\enddata
\tablecomments{Screen Orientation gives the direction of the long axis of the distribution of scattering images, which equals the direction of greatest change in electron density of the scintillation screen (and thus perpendicular to possible associated extended physical structures); Screen Velocity is measured parallel to the direction of Screen Orientation unless otherwise indicated and is relative to the Solar System barycentre; Galactic Rotation is the expected transverse motion of gas relative to the Solar System barycentre in the direction of the Screen Velocity; Predicted \HI Radial Velocity is the expected radial Local Standard of Rest velocity from Galactic rotation at the position of the screen.
  Errors given for the predicted \HI radial velocity and Galactic rotation only consider the uncertainty in the screen distance.
    The latter is dominated by uncertainties in fitting the scintillation for all but PSR J1141-6545, the only source for which no accurate distance is known from timing or parallax measurements.}
\tablenotetext{a}{Screen a of this pulsar transitioned from weak to strong scattering which was modelled as a change in the screen orientation with other values fixed.  Unlike other measurements, the screen velocity for screen a is given as the total magnitude of the screen transverse velocity rather than the projection of the velocity in the direction of the screen orientation.  The direction of the screen velocity is $101\pm 17^\circ$ so the Galactic rotation is given as the projection along this direction.  For full details see \citet{Sprenger2022}.}
\tablenotetext{b}{Associated with \HII region, Sh 2-27, which is embedded in the shell of the Local Bubble}
\end{deluxetable*}

\subsection{ISM Tracers}

We looked at several ISM tracers to find structures with similar distance and/or orientation as the scattering screen.
  Since many of our sources are at high Galactic latitude, however, most of the surveys we looked at were incomplete or had too diffuse or too low spatial resolution to identify filamentary structures.\footnote{For completeness, the maps we investigated but were not useful to find associations were the S-band Parkes All Sky Survey \citep[S-PASS,][]{Carretti2010} for polarized emission, the IRIS all-sky maps \citep{IRIS} for dust emission, and CO maps \citep{Dame2001}.}
  The exceptions to this were in \HI and H$\alpha$ maps.

For investigating \HI emission, we used the Galactic All-Sky Survey \citep[GASS,][]{GASS-DR1,GASS-DR2,GASS-Update}, Effelsberg-Bonn \HI Survey \citep[EBHIS,][]{EBHIS}, and Galactic Arecibo L-Band Feed Array \HI\ survey \citep[GALFA-HI,][]{GALFA-DR1,GALFA-DR2}.
In these surveys, sheet-like or filamentary structures are commonly seen, and we can compare these with screen orientations.

We used the composite all-sky map of \citet{Finkbeiner2003} to identify H$\alpha$ structures.  This map is composed of images from the Wisconsin H$\alpha$ Mapper Sky Survey \citep[WHAM-SS][]{WHAM}, the Virginia Tech Spectral-Line Survey \citep[VTSS,][]{VTSS}, and the Southern H-alpha Sky Survey Atlas \citep[SHASSA,][]{SHASSA}.
We found noticeable emission structures only near PSR~J0737$-$3039A and PSR~J1643$-$1224.

\subsection{Filament Analysis} \label{section:filament-analysis}

To identify filamentary structures and measure their orientation, we use the Rolling Hough Transform (RHT), developed by \citet{Clark2014}.
For each pixel in an image, the RHT counts the number of pixels within a window with diameter $D_w$ that are correlated with each other along given orientations.  The ``RHT Spectrum" at that pixel position is then the fraction of correlated pixels along an orientation in the window, with only those over a threshold $Z$ recorded.
The threshold and window size, combined with the pixel scale $\Delta x$ of a given survey, set a minimum filament size that can be identified, $ZD_w\Delta x$; we list relevant numbers for the different surveys in Table \ref{table:resolutions}.
For each of our pulsars, we look for filaments in the RHT spectrum at the sky position of the pulsar, and for the direction perpendicular to that of the scattering points.

\begin{figure*}
    \centering
    \includegraphics[width=0.8\textwidth]{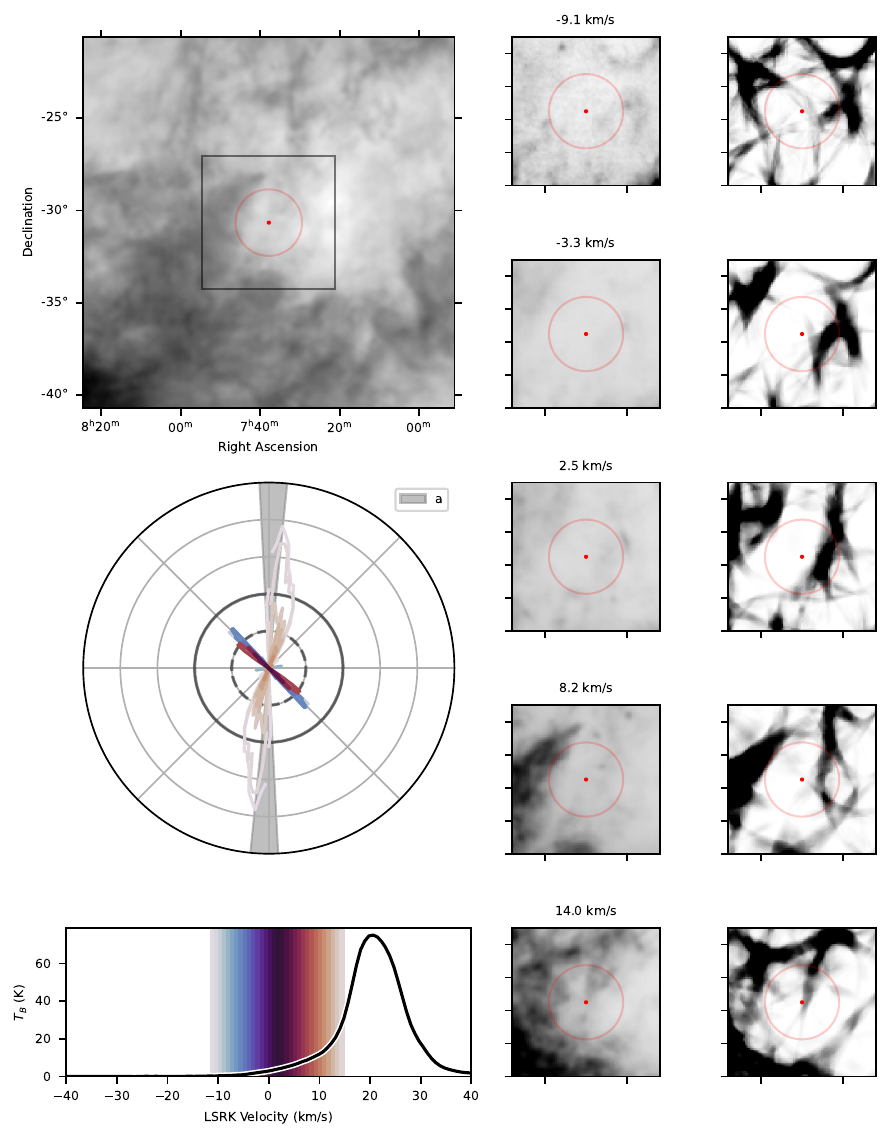}
    \caption{
      \HI filaments near PSR\,J0737$-$3039A. The velocity channels around $14{\rm\;km/s}$ show a filament well aligned with the orientation of the pulsar's scattering screen.
      {\em Left column, top:\/} \HI brightness temperature integrated over velocites between $-11.5$ and $14.8{\rm\;km/s}$ relative to the kinematic local standard of rest.
      The location of the pulsar is marked with a red dot, and the window for the Rolling Hough Transform analysis with a red open circle.
      {\em Left column, middle:\/} Rolling Hough Transform spectra at the position of the pulsar for a range of velocity channels, using colours from the panel below.
      The orientation of the pulsar's scattering screen is marked with a grey shaded region.
      {\em Left column, bottom:\/} \HI line profile towards the pulsar, with a colour bar overlaid to show the velocity channels used above.
      {\em Middle column:\/} \HI brightness temperature for different velocity channels (as labelled).  The region covered is indicated by the inset box in the top left panel.
      {\em Right column:\/} Corresponding Rolling Hough Transform backprojections.\\
      The complete figure set (10 images) is available in the online journal.}
    \label{fig:HIRHT}
\end{figure*}

\begin{deluxetable}{l c c c}
\tablecaption{\label{table:resolutions}  Rolling Hough Transform Parameters and Survey Resolutions}
\tablehead{  \\  & GASS & EBHIS & GALFA-HI}
\startdata
$\theta_{\mathrm{FWHM}}$ (arcmin)  & 16 & 10 & 4\\
$\Delta v_{\mathrm{LSRK}}$ (km/s)  & 0.824 & 1.288 & 0.184\tablenotemark{a} \\
$\Delta x$ (arcmin/pix) & 5 & 3.26 & 1\\
RHT window $D_w$ (pix) & 45 & 45 & 60\\
Minimum filament size (deg)  & 1.87 & 1.22 & 0.54
\enddata
\tablecomments{$\theta_{\mathrm{FWHM}}$ is the angular and $\Delta v_{\mathrm{LSRK}}$ the spectral resolution of the survey.
  A Rolling Hough Transform threshold $Z=0.5$ was used to calculate the minimum filament size.}
\tablenotetext{a}{For the Rolling Hough Transform analysis, the data were rebinned to 1.288 km/s.}
\end{deluxetable}

In Figure~\ref{fig:HIRHT}, we show the \HI emission data and associated RHT spectra for each pulsar.
For each of the sources, we used the highest resolution \HI survey available.
For three sources (PSR~J0337+1715, PSR~B0834+06, and PSR~B1133+16), data from both the GALFA-HI survey and EBHIS surveys were available.
We found that the two surveys agreed on alignments for these, although the higher resolution made the GALFA-HI data superior, with more filaments identified and narrower ranges of orientation angle.

The \HI data is resolved into velocity channels that give some information about the distance to the structures.
The conversion between velocity and distance is complicated by the complex velocity structure of the gas, but still allows us to separate very distant emission from the mostly nearby structures that we are interested in.
Hence, we include a velocity constraint in our requirements for alignment.

Note that because Galactic rotation does not change quickly with distance, each \HI velocity channel includes information from several hundreds of parsecs of gas, potentially with overlapping filaments of differing orientations.
  In contrast, scintillation screens occur at specific locations along the line-of-sight, so these are unlikely to correlate well with averaged values of alignment (and indeed, we did not find evidence of correlation with orientation from dust polarization).
  This complicates the definition of alignment, especially for lines of sight with multiple screens.

\subsection{Tests for Alignment}\label{section:stats-analysis}

To determine where a screen is aligned with a filament, we need to define what constitutes being aligned, as well as to estimate how likely it is that the alignment is due to chance.
We tried two methods, one where we set somewhat ad-hoc specific requirements, and another where we attempt to turn these constraints into a numerical score.
For each method, we quantify the statistical significance by simulating 2,000 random screens for each real screen, with random orientation and positions randomly chosen in $5^\circ \times 5^\circ$ squares of the sky centred on the corresponding pulsar.
A trial consists of one realization of a random screen for each real screen.\footnote{Our trials are limited in that they only compare against the null hypothesis of screen orientations being completely uncorrelated with those of filaments.
  We do not try to account for possible false negatives, i.e., the inability to observe true alignment due to, e.g., observational biases in the detection of screens and/or filamentary structures.
  It may become possible to quantify the false negative rate as more detailed observed three-dimensional filament distributions become available, or as magnetohydrodynamic simulations become realistic enough to identify both screens and filaments in them.}

Note that the two methods effectively test different questions, viz.,
whether an aligned filament is present, or whether the amount of
aligned emission is significant.  As we will see below, both methods
suggest that the screens are likely associated with filaments, but the
estimated probabilities of it being due by chance are quite different.

\subsubsection{Method 1: Specific Requirements for Alignment}

For our first method, we consider a filament aligned with a screen if:

\begin{enumerate}
\item Its velocity is within $\sim\!13{\rm\,km/s}$ of that expected from Galactic rotation (19 velocity channels for EBHIS and GALFA-HI data, 33 velocity channels for GASS data), i.e., offset by no more than roughly the speed of sound (for all pulsars with multiple screens, this selects essentially the same velocity range for all screens);
\item The difference between the screen orientation and a peak in the RHT spectrum is less than the error in the screen orientation added in quadrature to the error in the filament orientation.  We estimate the error in the filament orientation to be $15^\circ$ based on the average interquartile width of RHT spectra of 27$^\circ$ found in \citet{Clark2014};
\item At least 70\% of the pixels are aligned with the screen orientation in one velocity channel; and
\item At least 60\% of pixels are aligned with the screen orientation in each of four contiguous channels.
\end{enumerate}

Note that the last two conditions are similar to setting thresholds on the RHT, so this method is independent of choice of $Z$ as long as $Z \leq 0.6$. The likelihood of a screen classification as aligned arising by chance was then determined from the fraction of random screens within a $5 ^\circ \times 5^\circ$ region of the sky surrounding the real screens that were classified as aligned.
Similarly, the total likelihood of finding by chance as many screens aligned as observed, was determined from the number of trials where the number of aligned random screens were equal to or greater than the number of aligned real screens.

While this method is fairly straightforward, it is somewhat ad-hoc and sets hard cut-offs.
Furthermore, it is biased towards finding alignment by comparison with only the closest peak in the RHT spectrum (though this should not affect probabilities, as the trials share this bias).

\subsubsection{Method 2: Using an Alignment Score}

To create a quantitative score for whether a screen is aligned, we use a procedure inspired by the above method: we convolve the RHT spectrum with a filter in both orientation and velocity, and look for peaks at the screen orientation angle.
In orientation angle, to account for both the screen orientation measurement uncertainty $\delta\theta$ and our assumed systematic uncertainty of $15^\circ$ in the orientation of the filament relative to that of a screen, we filter using a  Gaussian with a width $\sigma = \sqrt{(15^\circ)^2 + (\delta \theta)^2}$.
For the velocity direction, to prioritize true filaments that extend over multiple velocity channels, we filter using a top hat with a width of four channels for the EBHIS and GALFA-HI data and six channels for the GASS data.
For each real and random screen, we then find the maximum value of the filtered RHT spectrum at the observed screen orientation and calculate a score $X$ of how many standard deviations above the mean of the filtered RHT spectrum the maximum is.

For each real screen, we estimate the probability $p_{\mrm{scr,}i}$ of its score arising by chance from the fraction of the random samples that has a higher value of $X$ than the real screen.
For all screens, we can then calculate a combined probability $p_{\mrm{tot}} = \prod_i p_{\mrm{scr,}i}$ (i.e., the joint probability), and use fraction of the trials with higher $p_{\rm tot}$ as an estimate of the likelihood that screens and filaments seem aligned by chance.

To get a sense of the uncertainties in our estimates, we repeat the above procedure for two values of the threshold: our main choice, $Z=0.6$, which aims to correspond to the threshold of method~1, and an extreme choice for comparison, $Z=0$, i.e., simply counting the number of aligned pixels with positive flux.

\section{Results for Individual Pulsars}\label{section:psr-idv}
\subsection{PSR J0737-3039A}

This project began by investigating possible associations of structures in the interstellar medium with the scattering screen towards PSR~J0737$-$3039A, a $23{\rm\;ms}$ pulsar in a binary with another, regular, $2.8{\rm\;s}$ pulsar, PSR J0737$-$3039B \citep{Lyne2004}.
The ability to time both pulsars (until Pulsar B precessed out of sight in 2008), short orbital period of $2.4{\rm\;hr}$, and eccentric orbit makes this system ideal for tests of general relativity.
Accordingly, this system has been studied very well \citep[e.g.,][]{Lyne2004,Kramer2006,Kramer2021a,Kramer2021b}.

Several studies have also used the scintillation of PSR~J0737$-$3039A to determine system and screen properties \citep{Ransom2004,Stinebring2005,Rickett2014}.
An interesting surprise was that the secondary spectra of this pulsar seem to show only one arc, indicating a single dominant screen \citep{Stinebring2005}, even though the system is very close to the Galactic plane ($\ell = 245^\circ, b = -4.5^\circ$), and the line-of-sight thus traverses gas with relatively large column densities and more complicated kinematics.

\citet[\citetalias{Rickett2014} hereafter]{Rickett2014} used orbital and annual variations in the scintillation timescale to determine the screen distance and orientation, but found two different solutions, one more and one less asisotropic.
To elucidate which one was better, we re-fit the scintillation timescales, choosing fit parameters somewhat better suited for constraining screen properties and using the updated pulsar timing ephemeris from \citet{Kramer2021b}.
Details of our fits are given in the Appendix, but here we simply note that the results are largely consistent with those of \citetalias{Rickett2014}, with the biggest difference being the adopted pulsar distance ($d_p$): \citetalias{Rickett2014} used $d_p=1.5{\rm\;kpc}$, based on the parallax measurement of \citet{Deller2009}, while we use $d_p=0.735{\rm\;kpc}$, the weighted average of parallax and timing parallaxes given by \citet{Kramer2021b}.
Since we find the same fractional distance, $s=1 - d_s/d_p=0.7$ (with $d_s$ the distance of the screen), our fit places the screen closer, at $d_s=260\pm50{\rm\;pc}$ compared to the $450{\rm\;pc}$ inferred by \citetalias{Rickett2014}.

We also found clearer evidence that the scattering screen is highly anisotropic, with its line of images nearly aligned with right ascension, at $\sim\!88\fdg6$ East of North.
We made fits for several possible values of the inclination but found that these had very little effect on the screen properties; in Table \ref{table:psr-lit-review}, we list the result for what formally is our best fit, with an inclination of $i=89\fdg35$.

% \subsubsection{Interstellar Medium Counterparts}\label{section:ism}

Comparing the screen orientation with \HI surveys, we find possible alignment with an \HI filament that spans velocities in the range of 14 to $27{\rm\;km/s}$; see Figure \ref{fig:HIRHT}.  The radial velocities of the possibly aligned filament somewhat exceed the expected radial velocity from Galactic rotation ($2.28{\rm\;km/s}$).
We note, however, that the tangential velocity of the scattering screen is also in excess of what is expected from Galactic rotation by a similar amount, $14{\rm\;km/s}$.
Despite the rather complex web of filaments in this section of the sky, the range of orientations identified with RHT is quite small.
Applying the RHT to the H$\alpha$ data did not identify filaments in front of PSR~J0737$-$3039A.

For this particular pulsar, the evidence for the presence of an aligned \HI filament is suggestive but in itself inconclusive: Method 1 finds similarly aligned filaments for 31.8\% of the random trials, while Method 2 shows that for only 1.2\% of the trials there is aligned emission equally or more significant than that for the pulsar for a threshold of 0.6 or a more similar 44.3\% for a threshold of 0.0.
The large discrepancy is likely due to this pulsar being close to the Galactic plane: while the line-of-sight to PSR J0737-3039A has only two clear filaments, many of the randomly chosen nearby lines of sight have multiple filament orientations.
Hence, they relatively easily fulfill the requirements for Method 1, which just looks for an aligned structure, while for Method 2, which also looks at the strength of of the aligned filament relative to others, those multi-filament trials are less likely to give a significant signal (at least if a cut-off is imposed).
As will become clear for the other pulsars, this large a discrepancy between the two methods is not common, but it nevertheless shows how strongly the significance of a result can depend on the choice of detection criteria.
Overall, we find the association suggestive.

One of the most prominent ISM structures along the line-of-sight to PSR J0737$-$3039A is the Gum Nebula. 
The Gum Nebula is a large, irregular ionized nebula with an average radius of $22.7\fdg\pm0\fdg1$ \citep{Purcell2015}.
It is likely an \HII region with the shell resulting from the winds of hot stars in its interior \citep{Purcell2015}.
The nebula and shell would have large density variations and thus likely favorable conditions for forming structures that could cause scintillation.

The distance to the Gum Nebula is uncertain, with estimates for its centre ranging from 350 pc to 500 pc \citep[e.g.][]{Woermann2001,Sushch2011,Purcell2015}.
\citet{Kramer2021b} argue that PSR~J0737$-$3039A is likely behind the Gum nebula, and suggest an extension of it, seen in emission near the pulsar's position, could be the source of the scintillation screen.
 However, the unknown distance to the Gum Nebula and its irregular shape, as well as the ambiguity in the distance to J0737-3039A make it difficult to determine if the screen distance is consistent with the Gum Nebula.  If the centre of the Gum Nebula is less than $\sim 400$ pc, or the pulsar is farther, the scintillation screen may be consistent with the front edge of the Gum Nebula.

The Gum Nebula's shell's expansion has been found to be slow and non-uniform, but the back face is expanding at $8.5{\rm\;km/s}$ while the front face is expanding at $14{\rm\;km/s}$ \citep{Woermann2001}.
This expansion would suggest a filamentary structure on the front face of the nebula would have a radial velocity of about $-11{\;\rm km/s}$ (after accounting for the motion of the centre of the nebula relative to Galactic rotation of 3.6 km/s used in \citet{Woermann2001}), which is inconsistent with the 14 to $27{\rm\;km/s}$ radial velocity of the potentially aligned filament.

While an association with the Gum Nebula may be possible, the screen distance is also similar to the distance of the Local Bubble wall estimated to be at 236 pc along the line-of-sight to J0737-3039A \citep{Pelgrims2020}. The Local Bubble is a low density, irregularly shaped cavity with radius of $\sim 100-300$ pc filled with hot ($\sim 10^6$ K), mostly ionized gas surrounded by a thick wall ($\sim 100$ pc) of dense, neutral gas.
The Local Bubble is thought to have formed from a series of supernova shocks 10 to 15 million years ago, with surviving stars now located in the Sco-Cen association \citep{Maiz-Appelaniz2001}.

Radial velocities of stars near the surface of the Local Bubble are around 5 to 9 km/s in excess of the Local Standard of Rest, and idealized modelling of the expansion of the Local Bubble gives a current expansion rate of $6.7^{+0.5}_{-0.4}$ km/s \citep{Zucker2022}.  This is somewhat smaller than the velocity of the filament but in the same direction.

In summary, we find a neutral hydrogen filament that is aligned with the screen.
There might be an association of the screen with either the Local Bubble or the Gum Nebula.
Both also have significant \HI components, with the Local Bubble wall defined by an increase in neutral gas and dust, and the Gum Nebula being surrounded by an \HI shell.
The distance to the scintillation screen and the velocity of the potentially aligned \HI filament favour the Local Bubble wall.

\subsection{PSR J0337+1715}
PSR~J0337$+$1715 is a millisecond pulsar in a hierarchical triple system.
The pulsar is in a 1.6 day orbit with a white dwarf within a 327 day orbit of another white dwarf.
The orbital parameters of this system are well known from pulsar timing and the effects of three-body interactions \citep{Voisin2020}.
The screen parameters for this system were fit from nine years of scintillation measurements by \citet{Gusinskaia_inprep}.

There is substantial \HI emission along the line-of-sight, but the orientation of the screen is not consistent with any identified \HI filament.
Indeed, while an aligned filament was found only for 5.3\% of our trials for nearby lines of sight, the maximum observed amount of aligned emission is less than the mean, and higher maxima are found for 96.9\% of the trials in method~2 (with $Z=0.6$).

\subsection{PSR J0437-4715}

PSR~J0437$-$4715 is a millisecond pulsar in a 5.76-day binary with a white dwarf.
It has a precisely measured distance that places it very near to us, at $156.79 \pm 0.25$ pc \citep{Reardon2016}.
Using 16 years of archival data, \citet{Reardon2020} were able to make precise measurements of two screens from arc curvature variations, with a primary screen at $89.8 \pm 0.4$ pc and a fainter, secondary screen at $124 \pm 3$ pc.
Two other arcs were identified in the data, implying this pulsar has at least four screens, but there was insufficient signal to measure the properties of these additional screens \citep{Reardon2020}.

Some filamentary structure is identified in front of this pulsar that appears to be aligned with both screens.
\HI emission in the direction of this pulsar is very weak (peaking at $\sim 1{\rm\;K}$ in 1.288 km/s channels), which could influence the reliability of identifying filaments.
However, the \HI morphology does not appear to be dominated by noise.

From our trials, the probability of finding a chance association for a single screen was quite large, at 37.3\%.
The likelihood that both screens would be aligned by chance was found to be 4.9\%, which looks more significant, but depends in part on the two observed screens having very similar orientation, while in the trials the orientations are randomized.
For the trials, we found that 7.6\% had stronger aligned emission than the nearer, primary screen, and 11.0\% of trials showed stronger aligned filaments than the farther, secondary screen.

\subsection{PSR J0613-0200}
PSR~J0613$-$0200 is a millisecond pulsar in a relativistic binary with a white dwarf companion in the Galactic plane \citep{Lorimer1995}.
From a seven year dataset, \citet{Main2020} fit for the orientation and distance to its screen using scintillation arcs, ignoring the contribution of orbital motion to the velocity of the system.
Increased scattering was observed in 2013 that could indicate a different screen.
The arc measurements in 2013 were not fit well either by a model that assumed a single screen for the full seven year dataset nor when fitted separately, resulting in the large error in the distance to this screen.
The best fits for both screens are shown in Table \ref{table:psr-lit-review}.

For both screens, we find possible aligned \HI filaments.
In this region of sky the likelihood to find at least one screen aligned by chance is 23.9\% and the likelihood of two screens being aligned by chance is 1.7\%.
The latter may exaggerate the case for aligned filaments, however, given that filaments are identified in front of this pulsar over a wide range of angles.
Indeed, we found that 50.7\% of trials showed more significant aligned flux than the 2013 screen and 22.5\% of trials showed more alignment flux than the 2014 and onwards screen.
Hence, the association, like for most screens taken on their own, is rather tentative.

\subsection{PSR J0636+5128}
PSR~J0636$+$5128 is a millisecond pulsar in a binary with a period of 1.6 hours \citep{Alam2021}.
The properties of the screen were determined by \citet{Liu2022,Liu2023} using the scintillation timescale and decorrelation bandwidth from 2.5 years of data and ignoring the effect of the orbital motion (which is small for this system, so did not significantly decrease the quality of the fit).

There is substantial \HI emission along the line-of-sight, with a clear filament identified between 2 and $10{\rm\; km/s}$ (exceeding the predicted radial velocity from Galactic rotation by 5 to $13{\rm\;km/s}$).
This filament, however, is not aligned with the screen but instead nearly  perpendicular.
The likelihood of finding a chance alignment along this line-of-sight is 20.9\%, and 65.3\% of trials showed more aligned emission than this screen.

\subsection{PSR B0834+06}

Using \ac{VLBI}, \citet{Brisken2010} mapped the positions of images on the sky for the screens of PSR~B0834$+$06.
They identified a roughly 40 mas line of images corresponding to the main arc of the secondary spectra, as well as cluster of images offset by about 10 mas from the pulsar position.
The cluster of images was suggested to be due to multiple scatterings of the light by the two screens \citep{Liu2016}, with the second screen terminating so that the pulsar was not seen through it.
This model, which was confirmed in detail by the phase retrieval analysis done by \citet{Zhu2023}, gives very accurate orientations.

There is a clear filament centred around 100$^\circ$ that is inconsistent with either screen.
In this region of the sky the likelihood of finding at least one aligned screen by chance is 11.0\%.
Despite the lack of a specific aligned filament, only in 9.8\% of the trials there was more aligned emission than what is seen for the main screen, i.e., the fainter filament visible in the RHT at the corresponding position angle is not entirely insignificant.
In contrast, 36.6\% of the trials had more aligned emission than was observed for the second screen.

\subsection{PSR B1133+16}

PSR~B1133$+$16 is relatively nearby, at $ 372\pm 2$ pc \citep{Deller2019} and has displayed up to six different scintillation arcs \citep{McKee2022}.
By tracing how those arcs changed with time over a 30 year baseline, \citet{McKee2022} found screen parameters for five screens.
There are large error bars in the fitted values as often not all arcs can be seen and it is difficult to identify which arc corresponds to which screen.
Although the relative brightness of the arcs vary with time, the scattering regions remains statistically similar over 30 years, placing a lower limit on the transverse width of the scattering region of $\sim\!4\arcsec$, corresponding to a size of $\sim 0.0035\,$pc at a distance of 186 pc.

%{\red Which arc is brightest alternates over timescales of a few weeks CITE TURNER if paper gets published}

For the purpose of our statistical analysis, we consider screens c, d, and e as one screen, since their orientations and expected radial velocities overlap.
These screens all show probable alignment with hydrogen filaments.
A filament is also found within the uncertainties of the orientation of screen f.
The weak \HI emission along this line-of-sight means that velocity channels with velocities above 5 km/s are dominated by noise and give unreliable but low percent orientations, below the 60\% threshold used to determine alignment.
The likelihood of chance alignment of at least one of three screens in this region of the sky is 10.2\%, and 0.12\% for at least two screens.
We found that 77.3\% of trials had stronger aligned emission than screen b (which was not considered aligned), 66.6\% had more than screen c/d/e, and 56.8\% had more than screen f.
Given the large error bars on the orientations of the screens, it is unsurprising that their alignment is not very significant.

\subsection{PSR J1141-6545}

PSR~J1141$-$6545 is a millisecond pulsar in a relativistic binary with a white dwarf companion.
Most of the orbital parameters of this pulsar are measured to high precision from pulsar timing.
The distance to the pulsar, however, is less well constrained, but \citet{Reardon2019} fit the distance to the pulsar as well as screen properties from the scintillation timescale.
Their best fit for the distance to the pulsar was $10^{+4}_{-3}\,$kpc with a screen at 7.20$^{0.30}_{-0.22}\,$kpc.
The fitting did not assume a one-dimensional screen, instead the screen was fit as an ellipse (using a method similar to that of \citealt{Rickett2014}).
The axial ratio of the ellipse was found to be $2.14 \pm 0.11$, so the screen may not be very filamentary or may need to be described by more than one linear screen.

No filaments were identified in front of this pulsar in the velocity range searched.
This is not surprising since the filaments at this distance are likely to be unresolved and hence the \HI emission is low.
Consequently, we do not include this pulsar in our statistical analyses.

\subsection{PSR B1508+55}

The secondary spectra of PSR~B1508$+$55 usually display a single arc with flat arclets, which was explained as a result of two-screen scattering by \cite{Sprenger2022}.
The screen creating the main arc is nearby, at a distance of $127.1\pm1.5$ pc and the second screen is located close to the pulsar, at a distance of $1940\pm123$ pc \citep{Sprenger2022}.
The morphology of the secondary spectra of this pulsar also appears to change, from a fuzzy arc to a thin arc with clear inverted arclets.  This change in morphology is modelled as being caused by a change from strong to weak scattering of the closer screen, accompanied by a change in orientation from $142.4 \pm 0.4^\circ$ to $129.8 \pm 2.0^\circ$ \citep{Sprenger2022}.
\cite{Sprenger2022} noted that the closer screen is 1.37 pc from the line-of-sight of an A2 star at a distance of $120\pm8$ pc, but found no candidate associations for the further screen.

A screen with orientation $142.4 \pm 0.4^\circ$ is consistent with being aligned with an \HI filament.
%{\red This orientation corresponds to the strong scattering regime of PSR~B1508$+$55's secondary spectra which may have higher column density than the weak scattering screen so is more likely to be observed.}
However, this filament's orientation of approximately $144^\circ$ is less consistent with the orientation of $121\fdg8\pm2\fdg0$ of the weak-scattering screen.
The probability of finding at least one screen is aligned by chance in this region of the sky is 27.5\%.
Both screens showed much stronger aligned emission than random trials, with only 9.9\% of trials having stronger aligned emission than the weak-scattering screen and 1.5\% of trials have more than the strong-scattering screen.

We do not find an alignment \HI filament for the distant screen, but this may be expected: it may be too far away to be resolved and thus too faint.
We thus exclude the distant screen from our statistical analysis.

PSR~B1508$+$55 is quite distant ($2.10^{+0.13}_{-0.14}\,$kpc) but has a very large proper motion (implying a velocity of $936^{+61}_{-64}$ km/s, \citealt{Chatterjee2009}).
Given that the scintillation properties of PSR~B1508$+$55 have changed as it crosses different ISM environments \citep{Sprenger2022}, one might worry about a mismatch in time.
The EBHIS data was collected between 2010 and 2011, and \citet{Sprenger2022} used data from 2020 to 2022.
Between 2011 and 2021 the pulsar would have moved $-0.74\arcsec$ in RA and $-0.62\arcsec$ in declination, i.e., much less than the pixel size of the EBHIS data.
We therefore do not expect to see changes in the \HI emission structure that would explain the changes in scintillation properties.

\subsection{PSR J1603$-$7202}

\citet{Walker2022} fit scintillation arc measurements for PSR~J1603$-$7202.
During the 12 years covered, PSR~J1603$-$7202 displayed a significant increase in dispersion measure accompanied by decreases in scintillation timescale and decorrelation bandwidth, which \citeauthor{Walker2022} characterize as an extreme scattering event.
The dispersion measure then plateaued before decreasing to pre-event levels. Motivated by these changes in the line-of-sight ISM environment during their data set, \cite{Walker2022} allow for fits to different screens in different epochs.
If only one screen is considered, their best fit orientation is $140\pm13^\circ$ from Galactic North, at a distance of 2.47 kpc.
For three screens, all had orientations consistent with $120^\circ$, but with distances between 2.52 kpc and 3.33 kpc.

In Table \ref{table:psr-lit-review}, we include only the single screen model.
Filaments identified with RHT are not consistent with orientations of either $140^\circ$ or $120^\circ$.
However, given the large distances to the screen(s) it is not obvious that filaments would be resolved, and hence we exclude this pulsar from our statistical analyses.

\subsection{PSR J1643$-$1224}

PSR~J1643$-$1224 is a millisecond pulsar in a 147 day orbit with a white dwarf companion, at a distance of $760\pm16\,$pc \citep{Desvignes2016}.
Using five years of scintillation arc measurements, \citet{Mall2022} modelled the scintillation screen(s) of PSR~J1643$-$1224.
The secondary spectra for this pulsar displayed a single arc that appears to alternate between a thin arc, indicative of anisotropic screen, and a more diffuse arc that could in principle be indicative of either an isotropic screen or of multiple anisotropic screens.
But scatter broadening of the pulsar image favours the solution with two anisotropic screens \citep{Ding2023}.

For comparison with \HI filaments, we use the fit for two anisotropic screens.
The distances of the two screens were $129\pm15$ and $340\pm9{\rm\,pc}$.
The distance to the closer screen is consistent with that of the \HII region, Sh 2-27, centred at $112\pm3\,$pc \citep{Ocker202O} with a diameter of 34 pc \citep{Harvey-Smith2011}.

Both screens have probable alignment with an \HI filament.
We also compared the orientation of screens with H$\alpha$ filaments and, although filamentary structure was identified in front of this pulsar, the filaments were not aligned with the screens.
The alignment with \HI and not H$\alpha$ could suggest that the changes in electron density responsible for the nearer screen do not arise in the \HII region itself, but rather at, e.g., the interface with neutral hydrogen surrounding it.
The likelihood that at least one of the screens is aligned by chance is 72.8\%, and that of both being aligned is 23.9\%.
We found that 15.0\% of trials had stronger aligned emission than the closer screen, while 85.1\% of trials had stronger aligned emission than the farther screen.

\subsection{PSR J1909$-$3744}

PSR~J1909$-$3744 is a very precisely timed millisecond pulsar due to its narrow pulse width, which has enabled a precise distance measurement, of $1.152\pm0.003{\rm\,kpc}$ \citep{Reardon2021}.
The pulsar is in a binary with white dwarf companion.
The scintillation screen of PSR~J1909$-$3744 was modelled using 13 years of scintillation arc measurements by \citet{Askew2023}.
They searched for nearby H$\alpha$ sources and hot stars but found no associations. Additionally, \citet{Ocker2024} identified a molecular cloud complex, Corona Australis, located along the line-of-sight but at a distance of 150 pc so it is unlikely that it is associated with the screen at a distance of 590 $\pm$ 50 pc \citep{Askew2023}.

We identify prominent filaments in front of the pulsar with orientations around $150^\circ$, inconsistent with the orientation inferred for the scintillation screen.
The likelihood of chance alignment for this line-of-sight is 28.2\% and 24.8\% of trials showed stronger aligned emission than this screen.

\section{Aggregate Results}\label{section:psr-agg}

\subsection{Filament Alignment}

We found that for six of the eleven pulsars analyzed there were possibly aligned \HI filaments for at least one of their screens, for a total of 12 of 22 screens with possible alignments (see the summary in Table \ref{table:stat-summary}).
All of the possible aligned screens were within 600 pc from Earth, and all but one within 340 pc.
There is an observational bias towards measuring the scintillation of less scattered, brighter pulsars which are typically nearby and therefore have nearby screens.
The \HI filament analysis is also biased towards nearby filaments since, for a given physical size, nearby filaments have larger angular size and are thus more easily identified.

We evaluated the significance of the probable alignment of the full sample of pulsars using the methods outlined in Section \ref{section:stats-analysis}.
We ignored screens that are greater than $1{\rm\;kpc}$ away and counted screens c, d, and e of PSR B1133+16 as one screen since they share the same orientation within their errors.
With these adjustments, we are left with 17 screens, for 10 of which we found an aligned filament.
Comparing this with our random trials, we found that only in 0.004\% of those 10 or more screens were classified as having aligned filaments; for the trials, the expected number of aligned screens was 3.02.
For the second method, which is based on the significance of the aligned flux, the evidence is less strong, but still highly suggestive: only 1.70\% of the random trials showed more significant aligned flux than the real data using a threshold $Z=0.6$ for the RHT.
If the threshold is ignored, and the full RHT spectrum with a threshold of $Z=0$ is used, then only 0.90\% of our trials had more aligned flux than the real data.
We thus conclude that the alignment of scintillation screens with \HI filaments seen for our sample is unlikely to be due entirely to chance.

\begin{deluxetable*}{l c  c r  r r  r r}[t]
\tabletypesize{\footnotesize}
\tablecaption{\label{table:stat-summary}  Summary of statistics}
\tablehead{
 & & & &\multicolumn{4}{c}{\dotfill Method 2\dotfill}\\
 & &\multicolumn{2}{c}{\dotfill Method 1\dotfill} & \multicolumn{2}{c}{\dotfill($Z = 0.6$)\dotfill} &\multicolumn{2}{c}{\dotfill($Z = 0.0$)\dotfill}\\
Pulsar & Screen & Aligned? &(\%) & X & (\%) & X & (\%)}
\startdata
J0337$+$1715  & a & N & 5.3 & $-0.4$ & 96.9 & $-0.2$ & 99.9\\
% \hline
J0437$-$4715  & a & Y & 20.8 & 4.8 & 7.6 & 2.5 & 5.9 \\
              & b & Y & 21.4 & 4.7 & 11.0 & 2.5 & 5.6\\
% \hline
J0613$-$0200& 2014+ & Y & 12.8 & 2.7 & 22.5 & 2.0 & 15.8\\
             & 2013 & Y & 12.8 & 1.4 & 50.7 & 1.4 & 56.6 \\
% \hline
J0636$+$5128  & a & N & 20.9 & $-0.3$ & 65.3 & 0.7 & 53.6\\
% \hline
J0737$-$3039A & a & Y & 31.8 & 5.5 & 1.2 & 1.6 & 44.3\\
% \hline
B0834$+$06    & a & N & 5.4 & 4.0 & 9.8 & 3.2 & 0.6\\ %\citet{Zhu_inprep}
              & b & N & 5.9 & 1.9 & 36.6 & 1.8 & 21.4 \\
% \hline
B1133$+$16    & b & N & 3.4 & 1.2 & 77.3 & 0.9 & 66.7\\
          & c/d/e & Y & 3.8 & 1.5 & 66.6 & 0.8 & 78.2\\
              & f & Y & 3.5 & 0.9 & 56.8 & 1.0 & 58.1 \\
% \hline
B1508$+$55  & a (weak) & N & 14.6 & 3.6 & 9.9 & 2.2 & 6.4\\
     & a (strong) & Y & 15.0 & 4.8 & 1.5 & 2.6 & 1.1 \\
% \hline
J1643$-$1224  & a & Y & 48.3 & 3.5 & 15.0 & 2.2 & 14.6\\
              & b & N & 48.4 & 0.6 & 85.1 & 0.6 & 93.7\\
% \hline
J1909$-$3744  & a & Y & 28.2 & 2.9 & 24.8 & 1.9 & 30.6\\
\hline
\multicolumn{2}{l}{Combined probability} & \multicolumn{2}{r}{0.004} &\multicolumn{2}{r}{1.70}&\multicolumn{2}{r}{0.90} \\
\enddata
\end{deluxetable*}

\subsection{Screen Velocities}

We compared the measured screen velocities with the expected velocity from Galactic rotation at the position of the screen (both values are listed in Table \ref{table:psr-lit-review}).
The majority of screens (14 of 22) have velocities within 20 km/s of the Galactic rotation.
Since the sound speed of the ISM is roughly 10 km/s \citep{Goldreich1995}, the differences are not inconsistent with what is expected from random motion and/or slowly expanding or contracting interstellar medium, especially taking into account that most of the sample of screens are at high Galactic latitudes ($>20^\circ$ out of the plane), where motions are less well explained with Galactic rotation models. Errors in Galactic rotation due to errors in the distance to the screen are generally small ($<$ 1 km/s) and do not affect these results.

Some exceptions to the screen velocities being comparable to the Galactic rotation velocities are the screens of PSR~J0437$-$4715 and PSR~B1133$+$16.
For PSR~J0437$-$4715, both screens are approximately 40 km/s in excess of the Galactic rotation.  For B1133+16, some of the best-fit screen velocities are even larger, at more than $70$ km/s, but they also have large error bars, of more than 50\%.
Indeed, while four of the five screens have best-fit screen velocities more than $20$ km/s in excess of Galactic rotation, only two (screens c and d) have velocities inconsistent with Galactic rotation within the 1$\sigma$ uncertainties.
Hence, while this hints at more rapidly moving ISM, the large uncertainties hinder any strong conclusions.
It would seem worth measuring those velocities more accurately, e.g., using a dedicated VLBI campaign rather than relying on annual variations in arc curvature.

\section{Conclusion} \label{section:Concusion}

Our results showed correlations between the orientations of scintillation screens and hydrogen filaments that are unlikely to be due entirely to chance.
Our estimates of the likelihood of alignment vary between different methods (0.004\% and 1.7\%), and neither method accounts for screens or filaments that may exist but are not observed.  However, both methods are suggestive of a real association.  The association with \HI filaments is appealing because they are ubiquitous so can easily explain why scintillation screens are very common.

If the correlation in alignment between scintillation screens and hydrogen filaments is physical, it implies there is correlation between alignment of \ac{ISM} structures across many orders of magnitude of spatial scale -- from the milli-arcsecond scales of scintillation to the degree scales of \HI emission.
Where there were overlapping \HI surveys of different scales, it appeared that filaments were aligned in both the lower and higher resolution maps.
However, higher resolution maps gave narrower orientations and more filaments.
Even at the smallest resolution of 4 arcmin, the filaments did not appear to be resolved.
This could impact the conclusions of alignment if substructure within filaments is not aligned with larger scale structure.

Scintillation requires fairly large gradients, $\gtrsim\!10{\rm\,cm^{-2}/cm}$ \citep{Pen2014}, in the column density of electrons, but this may not translate to large volumetric electron density variations.
Electron density gradients could be expected on the surfaces of cold neutral filamentary structures if they are embedded in warmer ISM, such that there is a transition from cold neutral to warm ionized phases creating a partially ionized skin on the filament.
We did not see a correlation with H$\alpha$ emission that would trace the warm ionized medium but this is, at least partially, due to insufficient spectral and angular resolution of the H$\alpha$ maps.

The correlations in orientation of scintillation screens and \HI filaments do not necessarily indicate a physical connection; instead, \HI filaments could be tracers of the conditions that create scintillation screens.
For instance, \HI filaments appear to trace large-scale magnetic field orientation \citep{Clark2014}, and it seems plausible that the same magnetic field has ionized media aligned with it that may cause the scintillation \citep[see for example][]{Gwinn2019}, or that the scintillation occurs in magnetic reconnection sheets \citep{Pen2014}.

Indeed, magnetohydrodynamic simulations suggest that the presence of magnetic fields may plausibly help align screens and filaments despite their very different spatial scales: in the simulations, anisotropies arise with long axes aligned with the direction of the mean field \citep[e.g.][]{Cho2000}. The degree of anisotropy is scale-dependent, with smaller spatial scales showing greater anisotropy \citep{Beattie2020}.
  If so, the screens would be tracing the local orientation of the mean magnetic field.

Independently, our results suggest it is worthwhile to try to measure more screen orientations, and in particular try to improve the precision, e.g., using VLBI (as so far done only for PSR~B0834+06 by \citealt{Brisken2010}).
Furthermore, higher-resolution \HI studies would show more clearly whether filaments are truly associated with the scintillation regions, or simply trace similarly oriented structures.

Deeper understanding of the ISM environments for scintillation will require connections with multiple ISM tracers, ideally with true distances.
While distances to ISM structures are generally not known, 3D ISM tomography is rapidly advancing, particularly in dust, allowing for detailed distance measurements of, e.g., the Local Bubble, as well as of more isolated \HI filaments using the correlations between dust and \HI.
Similarly, our understanding of the 3-dimensional structure of the Galactic magnetic field is rapidly improving, which may provide another indication of what causes the scintillating structures.

\section*{Acknowledgments}

We thank the anonymous referee for very constructive comments, which helped improve the focus and clarity of this work.
We also thank Barney Rickett for sharing the data on the double pulsar system, and for clarifications on his results; Antoine Marchal, Bryan Gaensler, and Peter Martin for helpful discussions about the interstellar medium;  Susan Clark for advice about the Rolling Hough Transform; Jing Luo for assistance in the usage of {\tt pint}; Vincent Pelgrims for help in the usage and understanding of his Local Bubble model; and the University of Toronto and broader scintillometry community for insights and support.

MHvK is supported by the Natural Sciences and Engineering Research Council of Canada (NSERC) via discovery and accelerator grants.
The Wisconsin H$\alpha$ Mapper and its H$\alpha$ Sky Survey have been funded primarily by the National Science Foundation. The facility was designed and built with the help of the University of Wisconsin Graduate School, Physical Sciences Lab, and Space Astronomy Lab. NOAO staff at Kitt Peak and Cerro Tololo provided on-site support for its remote operation.
The Virginia Tech Spectral-Line Survey (VTSS) is supported by the National Science Foundation. The Southern H-Alpha Sky Survey Atlas (SHASSA) is supported by the National Science Foundation.
This work has made use of S-band Polarisation All Sky Survey (S-PASS) data.
This research has made use of NASA’s Astrophysics Data System Bibliographic Services.

\software{Astropy \citep{astropy2013,astropy2018,astropy2022},
  MatPlotLib \citep{Matplotlib},
  NumPy \citep{numpy},
  SciPy \citep{Scipy2020},
  {\tt pint} \citep{pint-ascl,pint-paper},
  Rolling Hough Transform \citep{Clark2014}}

\bibliography{bibliography}

\appendix

\section{Properties of the Scintillation Screen of PSR\,J0737$-$3039A}

We make a new fit for the scintillation screen properties of PSR~J0737$-$3039A based on the scintillation timescale data and analysis of \citet{Rickett2014}, hereafter referred to as \citetalias{Rickett2014}.
This new fit was motivated by two factors: (1) the original work found a rather large scintillation screen velocity, especially for the most anisotropic case ($> 100$ km/s); and (2) the pulsar timing measurements of this pulsar were recently updated by \citet{Kramer2021b}, including a significantly different distance to the pulsar.
Below, we outline the details of the model we use, following the description and nomenclature of \citetalias{Rickett2014}, and present the results of the fits.

\subsection{Scintillation Model}

The scintillation timescale, $T_{\rm{ISS}}$, is the time it takes for the auto-correlation function of the source brightness to fall by a factor $e$.
Hence, the scintillation timescale depends on the spatial scale of phase changes on the screen, known as the mean diffractive scale ($s_0$), as well as the relative velocities of the source, observer, and screen.

The velocity of the line-of-sight can be decomposed into the barycentric velocity of the pulsar binary, $\mathbf{V}_{\mrm C}$ and the orbital velocity of the pulsar.
The barycentric velocity is given by,
\begin{equation}
    \mathbf{V}_{\mrm C}  = \mathbf{V}_{\mrm P} + \mathbf{V}_{\mrm E} s/(1-s) - \mathbf{V}_{\mrm {IS}} /(1-s),
\end{equation}
where $\mathbf{V}_{\mrm P}$ is the proper motion of the pulsar center of mass, $\mathbf{V}_{\mrm E}$ is the velocity of Earth, $\mathbf{V}_{\mrm {IS}}$ is the velocity of the screen and $s=1-(d_{\mrm{screen}}/d_{\mrm{psr}})$ is the fractional distance from the pulsar to the screen.
This velocity is related to the commonly used ``effective velocity'' as $\mathbf{V}_{\mrm{C}} + \mathbf{V}_{\mrm{P,orbital}}= \mathbf{V}_{\mrm{eff}}/(1-s)$.

\begin{figure}
    \centering
    \includegraphics[width = 0.5\textwidth]{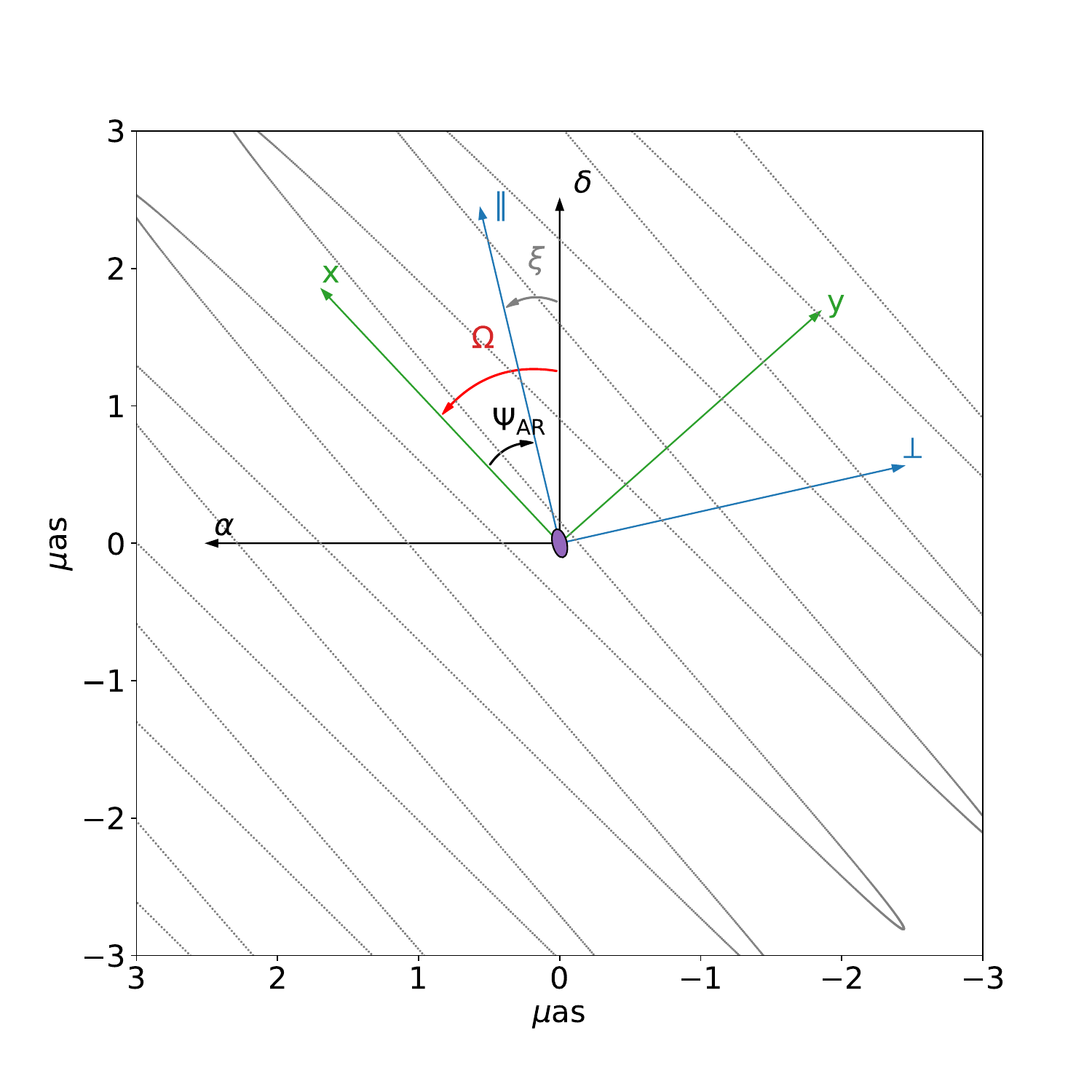}
    \caption{Schematic showing the coordinate systems used.
      The $x-y$ plane is oriented with $x$ along the line of nodes of the orbit.
      It has its origin at the equatorial coordinates $\alpha$, $\delta$ of the pulsar, but is rotated East of North (around equatorial coordinates ) by the longitude of ascending node, $\Omega$.
      The direction parallel to the semi-major axis of the screen resolution is rotated counterclockwise from the $x$ axis by the angle, $\psi_{\mrm{AR}}$.
      The orientation of the screen as seen from Earth, $\xi$, is measured East of North, again defined as that of the long, ``parallel'' axis.
      The screen resolution element is shown in purple for an axial ratio of 2.06, corresponding to an anisotropy parameter of $R=0.62$.
      It is to scale for an observation at $820{\rm\;MHz}$ and a distance of $262{\rm\;pc}$ (using the fractional distance to the screen of $s=0.643$ and a pulsar distance of $735{\rm\;pc}$).
      For comparison, the projected path of the pulsar over eight orbits is shown for $i=89\fdg35$, using grey dots, each separated by ten seconds.
      Note that it is the parallel direction along which to expect any filamentary structure, but it is the perpendicular direction along which the scintillation timescale changes fastest, and for which the screen velocity can be best constrained.
    }
    \label{fig:geometry}
\end{figure}

The orbital motion is most easily calculated in the reference frame of the pulsar orbit, defined such that $\hat{z}$ points from the Sun to the barycentre of the orbit, $\hat{x}$ is aligned with the line of nodes of the pulsar orbit, and $\hat{y}=\hat{z} \times \hat{x}$.
The transformation from equatorial coordinates to the pulsar orbital frame is done by rotating the velocities by the longitude of ascending node, $\Omega$, which is a free parameter.
The geometry of the model is shown in Figure \ref{fig:geometry}.
In x-y coordinates, the average orbital velocity is given as ($ V_{\mrm o} e\: {\sin}\:\omega,  V_{\mrm o}e \: {\cos}\: \omega \:{\cos}\: i)$, where $\omega$ is the argument of periastron.  $V_o$ is the average orbital speed, given as $V_o = 2 \pi a/(P_{\mrm b} \sqrt{1-e^2})$ where $a$ is the semimajor axis of the orbit, $P_b$ is the orbital period, and $e$ is the eccentricity of the orbit.

To determine screen parameters from the orbital harmonic coefficients, a single, potentially anisotropic scattering screen is assumed.  The anisotropy of the screen is described by its axial ratio, $A_{\mrm R}$, and the orientation of its major axis, $\psi_{\mrm{AR}}$.  Since $A_{\mrm R}$ is unbounded, \citetalias{Rickett2014} define a variable $R = (A^2_{\mrm R} - 1)/(A^2_{\mrm R} +1)$ which ranges from 0 (isotropic) to 1 (linear).  These parameters are contained within the following variables:

\begin{align}
    a &= \left[1-R \:{\cos} (2\psi_{\mrm{AR}})\right]/\sqrt{1-R^2},\nonumber\\
    b &= \left[1+R \:{\cos} (2\psi_{\mrm{AR}})\right]/\sqrt{1-R^2},\nonumber\\
    c &= -2R\:{\sin}(2\psi_{\mrm{AR}})/\sqrt{1-R^2}.
\end{align}

The scintillation timescale will depend on the orbital angle $\phi$ (which, like \citetalias{Rickett2014}, we define relative to the line of nodes, as $\phi=\theta+\omega$, with $\theta$ the true anomaly).
This dependence can be fitted by five orbital harmonic coefficients ($K_0,K_S,K_C,K_{S2},K_{C2}$) as
\begin{equation}
    T_{\rm{ISS}} (\phi) = K_0 + K_S \sin(\phi) + K_C \cos (\phi)
    + K_{S2} \sin(2\phi) + K_{C2} \cos (2\phi).
\end{equation}
The orbital harmonic coefficients are defined in terms of the physical parameters as,
\begin{equation}
    \begin{split}
      K_0 = & [0.5\: V_{\mrm o}^2 (a + b\: {\cos}^2\: i)
              + a(V_{\mrm C_x} - V_{\mrm o} e\: {\sin}\:\omega)^2
              + b(V_{\mrm C_y} + V_{\mrm o}e \: {\cos}\: \omega \:{\cos}\: i)^2\\
            & + c(V_{\mrm C_x} - V_{\mrm o} e\: {\sin}\: \omega )(V_{\mrm C_y} + V_{\mrm o} e \: {\cos}\: \omega \:{\cos}\: i)]/s_{\mrm p}^2 \\
      K_{\mrm S} =& -V_{\mrm o}[2a(V_{\mrm C_x}-V_{\mrm o}e\: {\sin}\: \omega)
                  + c(V_{\mrm C_y} + V_{\mrm o} e \: {\cos}\: \omega \:{\cos}\: i)]/s_{\mrm p}^2\\
      K_{\mrm C} = & V_{\mrm o}\: {\cos}\: i [c(V_{\mrm C_x} - V_{\mrm o} e\: {\sin}\: \omega)
                     + 2b(V_{\mrm C_y} + V_{\mrm o} e \: {\cos}\: \omega \:{\cos}\: i)]/s_{\mrm p}^2\\
    K_{\mrm{S2}} = & -0.5 c V_{\mrm o}^2\: {\cos}\: i /s_p^2 \\
    K_{\mrm{C2}} = &\; 0.5 V_{\mrm o}^2 (-a + b\: {\cos}^2 i)/s_{\mrm p}^2.\\
    \end{split}
\end{equation}
Here, the size of the resolution element of the screen, $s_{\mrm p}$, is related to the mean diffractive scale, $s_0$ by  $s_{\mrm p} = s_0/(1-s)$.

Since the mean diffractive scale is frequency dependent, scintillation timescales of data from observations in different frequency bands cannot be directly compared.
In order to combine data, \citetalias{Rickett2014} used normalized orbital harmonic coefficients ($k_0 = K_0/K_{\mrm {C2}}$, $k_{\mrm s} = K_{\mrm S}/K_{\mrm {C2}}$, etc.), which are independent of the mean diffractive scale.

In our fits, we depart from the fitting method used in \citetalias{Rickett2014} in two ways, to minimize the effect of covariances on the properties of the screen.
First, we fit directly for the angle $\xi$ of the scintillation screen on the sky (measured East from North, and constrained to fall in the range 0 to $180^\circ$), making the orientation $\Omega=\xi+\psi_{\rm AR}$ of the pulsar orbit a derived parameter (rather than vice versa).
This is because orbital motion directly constrains relative orientations between the screen and the orbits: $\psi_{\rm AR}$ for the pulsar, and $\xi$ for the Earth.
Indeed, in the fits from \citetalias{Rickett2014}, one sees that $\Omega$ and $\psi_{\rm AR}$ are strongly correlated, while their difference $\Omega - \psi_{\rm{AR}} = \xi$ is well constrained.

Second, we represent the velocity of the screen in directions parallel and perpendicular to the screen's semi-major axis rather than along the orbital $x$ and $y$ coordinates as done in \citetalias{Rickett2014}.
This is because in the limit that $R \rightarrow 1$ (i.e. a one-dimensional screen) the scintillation timescale is completely independent of motion parallel to the screen orientation, and hence the velocity along this direction becomes indeterminate.
By using velocity components in the screen coordinates, this uncertainty is directly captured by the fits, while in any other coordinate system, the velocities and uncertainties in both directions would be affected by it.

\subsection{Data and Fitting}

We fit for physical parameters of the pulsar and scattering screen using the normalized orbital harmonic coefficients presented in \citetalias{Rickett2014}.
These coefficients were measured from 17 observations taken between December 2003 and July 2005, using the 100 m Robert C. Byrd Green Bank Telescope at either 1850 MHz with a bandwidth of 800 MHz or 820 MHz with a bandwidth of 50 MHz.
For details about the observations and data reduction, see \citetalias{Rickett2014}.

For the velocity of the Earth, $\mathbf{V}_{\mrm E}$, we used the JPL DE401 ephemeris \citep{Park2021}.
The Earth velocity was calculated only for the beginning of the observation since the change in Earth velocity during an observation is very small.
The interstellar plasma velocity, $\mathbf{V}_{\mrm {IS}}$, and the fractional distance to the screen, $s$, were assumed to be constant between all epochs.
Orbital phases and scales for pulsar A were inferred using \textsc{pint} \citep{pint-ascl,pint-paper}, using the parameter file from \citet{Kramer2006} for comparison with R14, and that from \citet{Kramer2021b} for the results presented here.
Measurements of Shapiro delay constrain the value of $\sin i$ but are degenerate with the sign of $\cos i$.
Like \citetalias{Rickett2014}, we fit the scintillation timescales with the inclination fixed for both signs of $\cos i$ as well as fitting $i$ as a free parameter.
The inclinations used were $i = 88.7^{\circ}$ or $91.3^{\circ}$ from \citet{Kramer2006}, and $i = 89.35^{\circ}$ or $90.65^{\circ}$ from \citet{Kramer2021b}.

\subsection{Results} \label{section:Results}

\begin{deluxetable}{l c c c c c c}
\centering
\caption{Fitted Parameter, with 1$\sigma$ credible intervals\label{table:new-min-comparison}}
\tablehead{ & \multicolumn{2}{c}{\dotfill $\cos i > 0 $ \dotfill} & \multicolumn{2}{c}{\dotfill $\cos i < 0 $ \dotfill} & \multicolumn{2}{c}{\dotfill $i$ free \dotfill} \\
Parameter & $V_{\rm{IS},\parallel}$ free & $V_{\rm{IS},\parallel} = 0 $ & $V_{\rm{IS},\parallel}$ free & $V_{\rm{IS},\parallel} = 0$ & $V_{\rm{IS},\parallel}$ free & $V_{\rm{IS},\parallel}=0$}
\startdata
i (deg) & 89.35 & 89.35 & 90.65 & 90.65 & 87.8 $\pm$ 1.2 & 88.85 $\pm$ 0.54 \\
s & 0.646 $\pm$ 0.040 & 0.665 $\pm$ 0.031 & 0.614 $\pm$ 0.047 & 0.613 $\pm$ 0.045 & 0.708 $\pm$ 0.013 & 0.695 $\pm$ 0.023\\
$\xi$ (deg) \tablenotemark{a} & 178.6 $\pm$ 4.3 & 11.9 $\pm$ 4.7 & 174.2 $\pm$ 3.0 & 167.6 $\pm$ 3.9 & 176.3 $\pm$ 3.7 & 15.8 $\pm$ 5.0\\
R & 0.87 $\pm$ 0.11 & 0.63 $\pm$ 0.14 & 0.969 $\pm$ 0.057 & 0.840 $\pm$ 0.083 & 0.76 $\pm$ 0.17 & 0.44 $\pm$ 0.19\\
$\psi_{\rm{AR}}$ (deg) & 43 $\pm$ 10 & 34.9 $\pm$ 7.0 & 141.5 $\pm$ 8.9 & 148.7 $\pm$ 7.0 & 73 $\pm$ 13 & 43.6 $\pm$ 9.2\\
$V_{\rm{IS},\perp}$ (km/s) & -25.6 $\pm$ 1.5 & 23.7 $\pm$ 1.3 & -23.6 $\pm$ 1.9 & -21.9 $\pm$ 1.3 & -25.3 $\pm$ 1.7 & 25.3 $\pm$ 1.4\\
$V_{\rm{IS},\parallel}$ (km/s) & -46 $\pm$ 36 & 0 & 100 $\pm$ 120 & 0 & -47 $\pm$ 28 & 0\\
$N_{\rm{dof}}$ & 58 & 59 & 58 & 59 & 57 & 58\\
$\chi_\nu^2$ & 1.88 & 1.85 & 2.15 & 2.30 & 1.81 & 1.95 \\
$d_s$ (pc) \tablenotemark{b} & 260 $\pm$ 51 & 246 $\pm$ 23 & 284 $\pm$ 55 & 284 $\pm$ 33 & 215 $\pm$ 18 & 224 $\pm$ 17\\
$\Omega$ (deg) & 42 $\pm$ 11 & 47 $\pm$ 12 & 136 $\pm$ 12 & 136 $\pm$ 11 & 69 $\pm$ 10 & 59 $\pm$ 14\\
\enddata
\tablenotetext{a}{Screen orientation measured east from declination axis}
\tablenotetext{b}{Screen distance uses pulsar distance $d_p = 735 \pm 60$ pc}
\end{deluxetable}

\begin{figure}
    \centering
    \includegraphics[width = 0.5\textwidth]{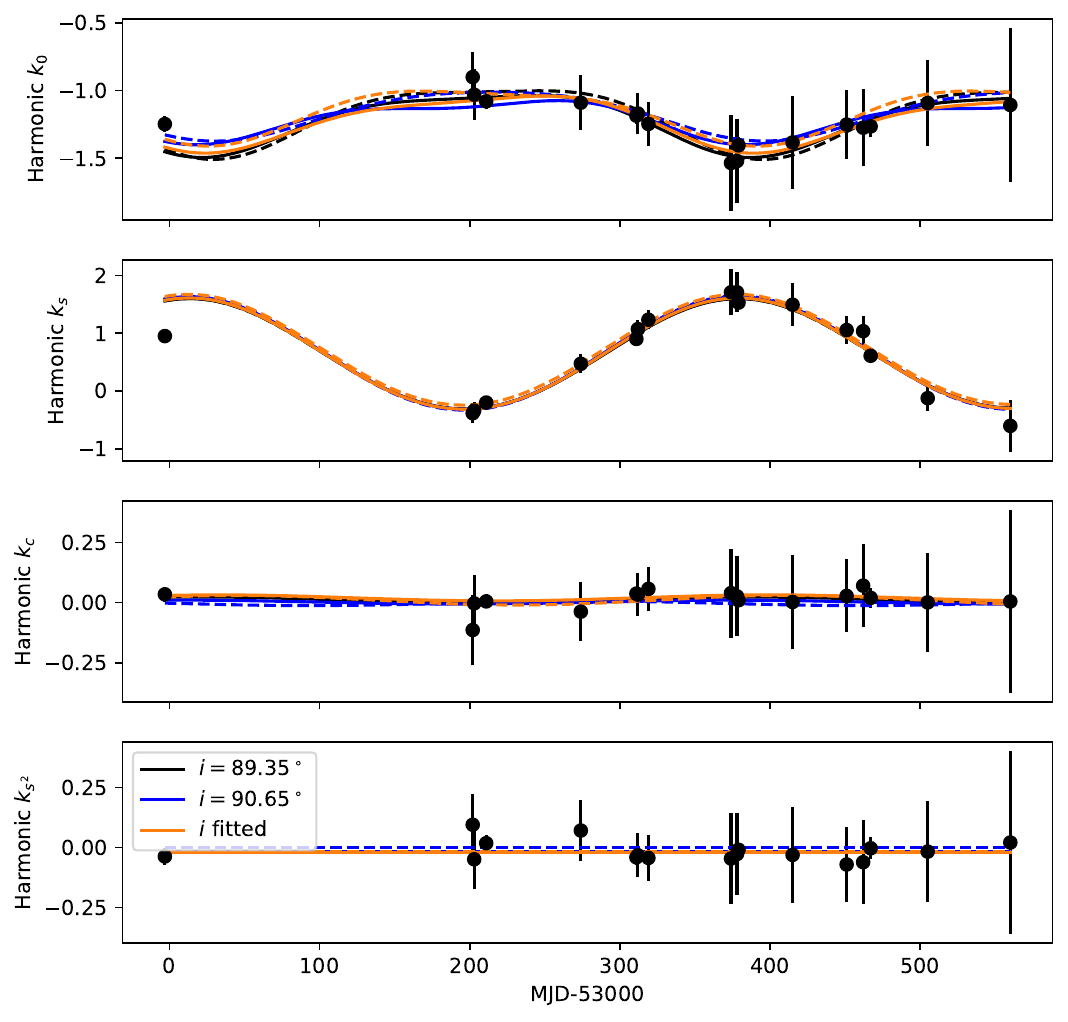}
    \caption{Normalized orbital harmonic coefficients and corresponding fits for different values of the orbital inclination, $i$, using the timing solutions from \citet{Kramer2021b}.
        Measured values from \citet{Rickett2014} are shown with black circles.  Solid lines indicate best fits where parallel screen velocity ($V_{\textrm{IS},\parallel}$) is free, while dashed lines indicate where $V_{\textrm{IS},\parallel}$ = 0 is fixed.
        One sees that for the cases where the inclination is fixed to the values allowed by the timing solution, the data are not sensitive to the parallel velocity.
        Note that first point is missed by most of our fits.
        Possible reasons for why this point deviates so strongly are discussed by \citet{Rickett2014}.
        It has little effect, however, on the inferred distance and orientation of the screen
        (as is clear from the fact that our fits match the other points rather well).
    }
    \label{fig:normk-fit}
\end{figure}

To test our fitting routine, we first used parameters like in \citetalias{Rickett2014}.
We found consistent results, except that our calculated reduced chi-squared values ($\chi_\nu^2$) differed somewhat, for reasons we do not quite understand but possibly simply related to having somewhat different codes and ephemerides.
For all further fits, we used the updated pulsar timing values from \citet{Kramer2021b}.
These give very similar results, which is not unexpected, since for our purposes, all parameter changes except for the distance are relatively small.

One of the biggest changes in the new timing parameters for PSR~J0737$-$3039A is the distance to the pulsar.
In our fit setup, the only effect of changing the pulsar distance is that it changes the systemic velocity of the pulsar binary, $\mathbf{V}_{\mrm P}$.
R14 used $V_{\rm{P}\alpha} = -17.8{\rm\;km/s}$ and $V_{\rm{P}\delta} = 11.6{\rm\;km/s}$ based on the distance of $1.15^{+0.22}_{-0.16}{\rm\;kpc}$ determined by \citet{Deller2009}.
More recent astrometry yields a distance of $770\pm70{\rm\;pc}$, while the parallax inferred from timing implies $465^{+134}_{-85}{\rm\;pc}$ \citep{Kramer2021b}.
As in \citet{Kramer2021b}, we adopted a weighted mean of those distances, of $735\pm60{\rm\;pc}$, and combine that with the the proper motion of $(-2.567, 2.082){\rm\;mas/yr}$ to calculate the velocities.

To see whether we could constrain the distance through scintillation, we tried separate fits with each of these three distances.
We found that we could not; all that changes is the fitted screen velocity (proportionally to distance).
This is not surprising, since constraining the distance requires constraining the pulsar angular orbital velocity, but for a nearly edge-on orbit and a nearly one-dimensional screen, this is highly degenerate with the angle $\psi_{\rm AR}$ between the screen and the orbit\footnote{For an overview, see \url{https://screens.readthedocs.io/en/latest/\#modelling-scintillation-velocities}.}, which we cannot constrain well at all.

Apart from the distance, the fits yield consistent results for all
other properties of the screen: a fractional distance from the pulsar $s\simeq0.7$, a high anistropy of $R\simeq0.9$, a roughly N-S orientation, and a perpendicular velocity of $v_\perp\simeq-25{\rm\;km/s}$.
These values consistent with those of \citetalias{Rickett2014}.

In all cases, the nominal parallel screen velocity is rather large but also is much more uncertain than the perpendicular velocity, as expected for anisotropic screens.
To test whether a large parallel screen velocity is required to obtain a good fit, we set $V_{\rm{IS},\parallel} = 0$ km/s and repeat the analysis.
The resulting fits are shown in Figure \ref{fig:normk-fit}.

Using the Akaike Information Criterion (AIC), we find that large velocities are not required: fits with $V_{\rm{IS},\parallel} = 0$ km/s are indeed worse but not by much, with the AIC increased from 121 to only 125 in the $i=89.35^\circ$ case (and from 137 to 146 for $i=90.65^\circ$, and from 117 to 125 with $i$ left free).
The fit also gives somewhat different axial ratios (systematically lower, between 0.44 and 0.84) and screen orientations (between 167.6$^\circ$ and 195.8$^\circ$), but only very slight changes in the perpendicular velocity (between 21.9 km/s and 25.3 km/s) and fractional screen distance (between 0.613 and 0.695).
The fact that there is some sensitivity to the parallel velocity suggests the screen is not fully linear (or that there is a contribution from a second linear screen).

We conclude that the dominant screen is anisotropic, and that our parameters of interest, the screen distance, orientation, and perpendicular velocity, are well constrained.

\clearpage

\begin{figure}
    \centering
    \label{fig:J0337}
    \includegraphics[width=0.8\textwidth]{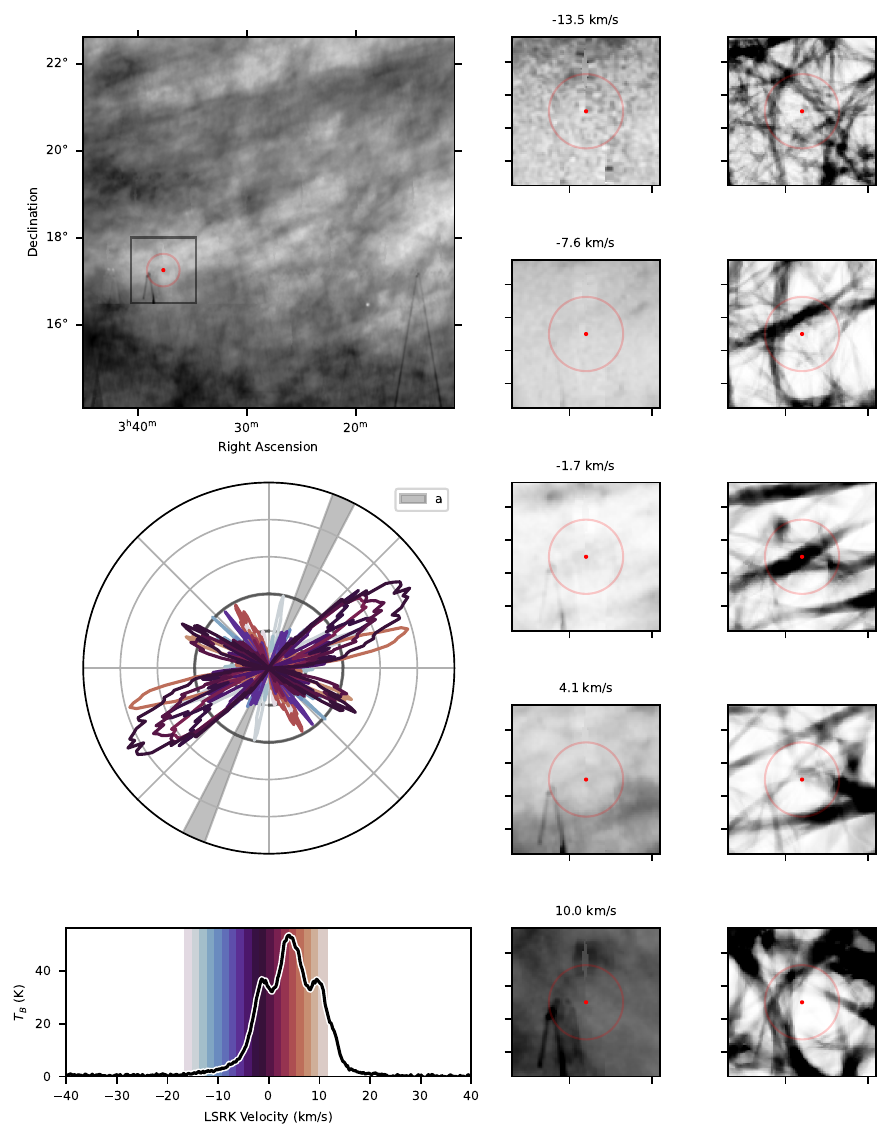}
    \caption{\HI filaments near PSR\,J0337+1715. The identified \HI filaments are not aligned with the screen.
      {\em Left column, top:\/} \HI brightness temperature integrated over velocites between $-16.5$ and $11.5{\rm\;km/s}$ relative to the kinematic local standard of rest.
      The location of the pulsar is marked with a red dot, and the window for the Rolling Hough Transform analysis with a red open circle.
      {\em Left column, middle:\/} Rolling Hough Transform spectra at the position of the pulsar for a range of velocity channels, using colours from the panel below.
      The orientation of the pulsar's scattering screen is marked with a grey shaded region.
      {\em Left column, bottom:\/} \HI line profile towards the pulsar, with a colour bar overlaid to show the velocity channels used above.
      {\em Middle column:\/} \HI brightness temperature for different velocity channels (as labelled).  The region covered is indicated by the inset box in the top left panel.
      {\em Right column:\/} Corresponding Rolling Hough Transform backprojections.\\}
\end{figure}

\begin{figure}
    \centering
    \label{fig:J0437}
    \includegraphics[width=0.8\textwidth]{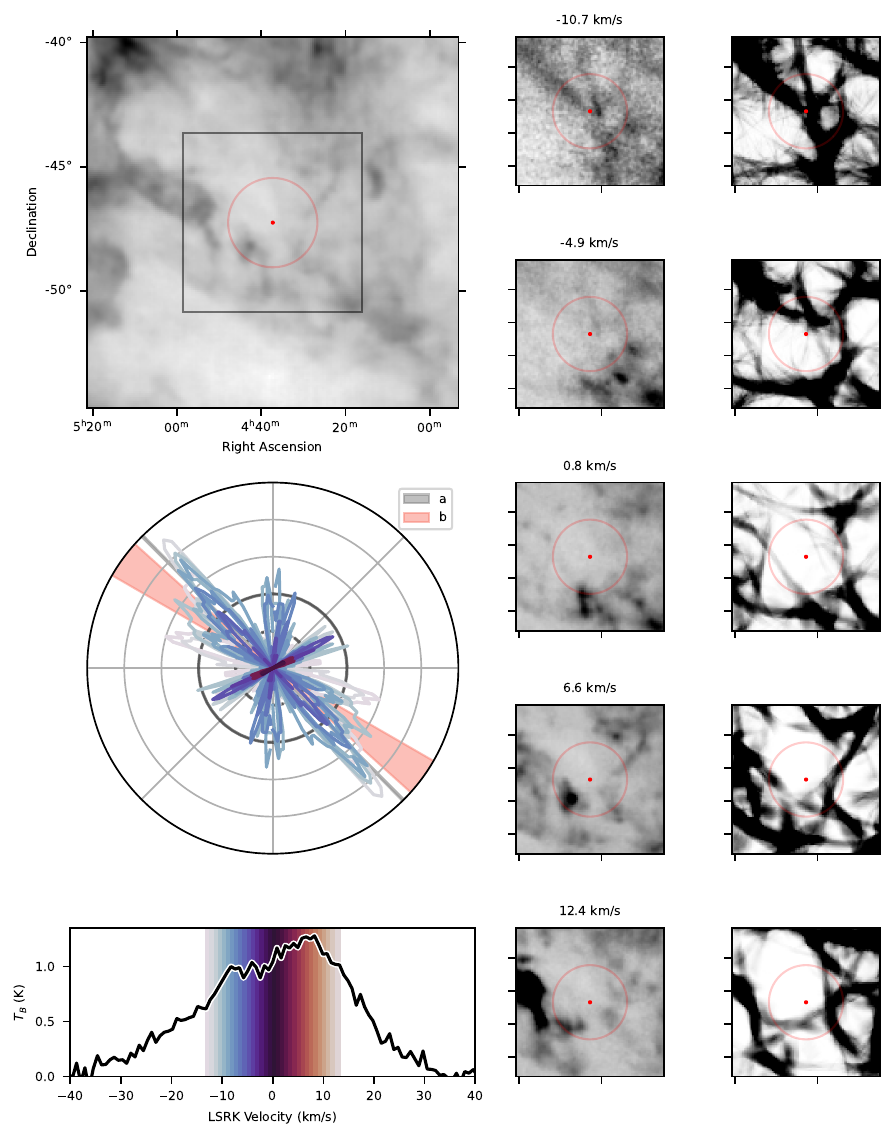}
    \caption{\HI filaments near PSR\,J0437$-$4715. There is probable alignment of \HI filaments with both screens.
      {\em Left column, top:\/} \HI brightness temperature integrated over velocites between $-13.2$ and $13.2{\rm\;km/s}$ relative to the kinematic local standard of rest.
      The location of the pulsar is marked with a red dot, and the window for the Rolling Hough Transform analysis with a red open circle.
      {\em Left column, middle:\/} Rolling Hough Transform spectra at the position of the pulsar for a range of velocity channels, using colours from the panel below.
      The orientation of the pulsar's scattering screen is marked with a grey shaded region.
      {\em Left column, bottom:\/} \HI line profile towards the pulsar, with a colour bar overlaid to show the velocity channels used above.
      {\em Middle column:\/} \HI brightness temperature for different velocity channels (as labelled).  The region covered is indicated by the inset box in the top left panel.
      {\em Right column:\/} Corresponding Rolling Hough Transform backprojections.\\}
\end{figure}

\begin{figure}
    \centering
    \label{fig:J0613}
    \includegraphics[width=0.8\textwidth]{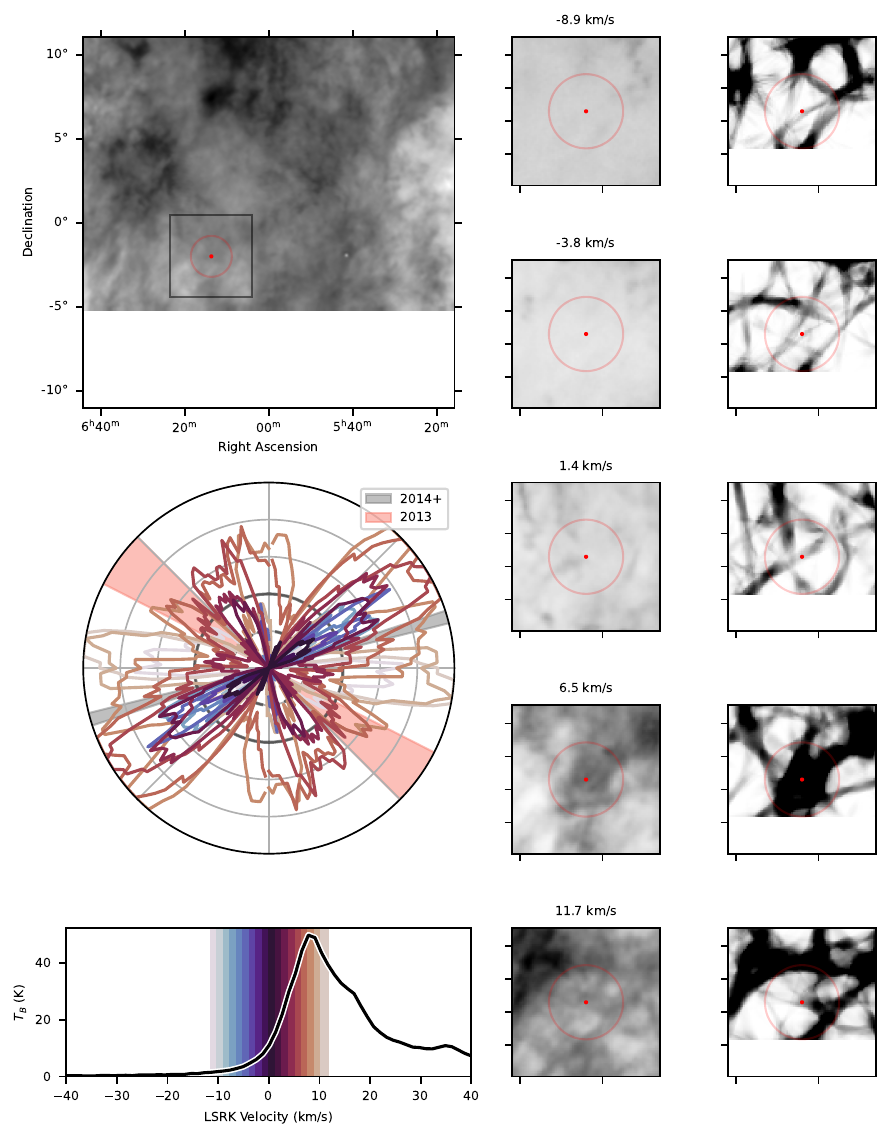}
    \caption{\HI filaments near PSR\,J0613$-$0200. There is probable alignment of \HI filaments with both screens.
      {\em Left column, top:\/} \HI brightness temperature integrated over velocites between $-11.5$ and $11.7{\rm\;km/s}$ relative to the kinematic local standard of rest.
      The location of the pulsar is marked with a red dot, and the window for the Rolling Hough Transform analysis with a red open circle.
      {\em Left column, middle:\/} Rolling Hough Transform spectra at the position of the pulsar for a range of velocity channels, using colours from the panel below.
      The orientation of the pulsar's scattering screen is marked with a grey shaded region.
      {\em Left column, bottom:\/} \HI line profile towards the pulsar, with a colour bar overlaid to show the velocity channels used above.
      {\em Middle column:\/} \HI brightness temperature for different velocity channels (as labelled).  The region covered is indicated by the inset box in the top left panel.
      {\em Right column:\/} Corresponding Rolling Hough Transform backprojections.\\
      Note there is a cutoff in the \HI brightness temperature map due to the declination cutoff of -5$^\circ$ for the EBHIS survey and a similar cutoff in the Rolling Hough Transform backprojections corresponding to one Rolling Hough Transform window from the edge of the \HI data.
      This did not affect the filament measurements at the position of the pulsar.
      }
\end{figure}

\begin{figure}
    \centering
    \label{fig:J0636}
    \includegraphics[width=0.8\textwidth]{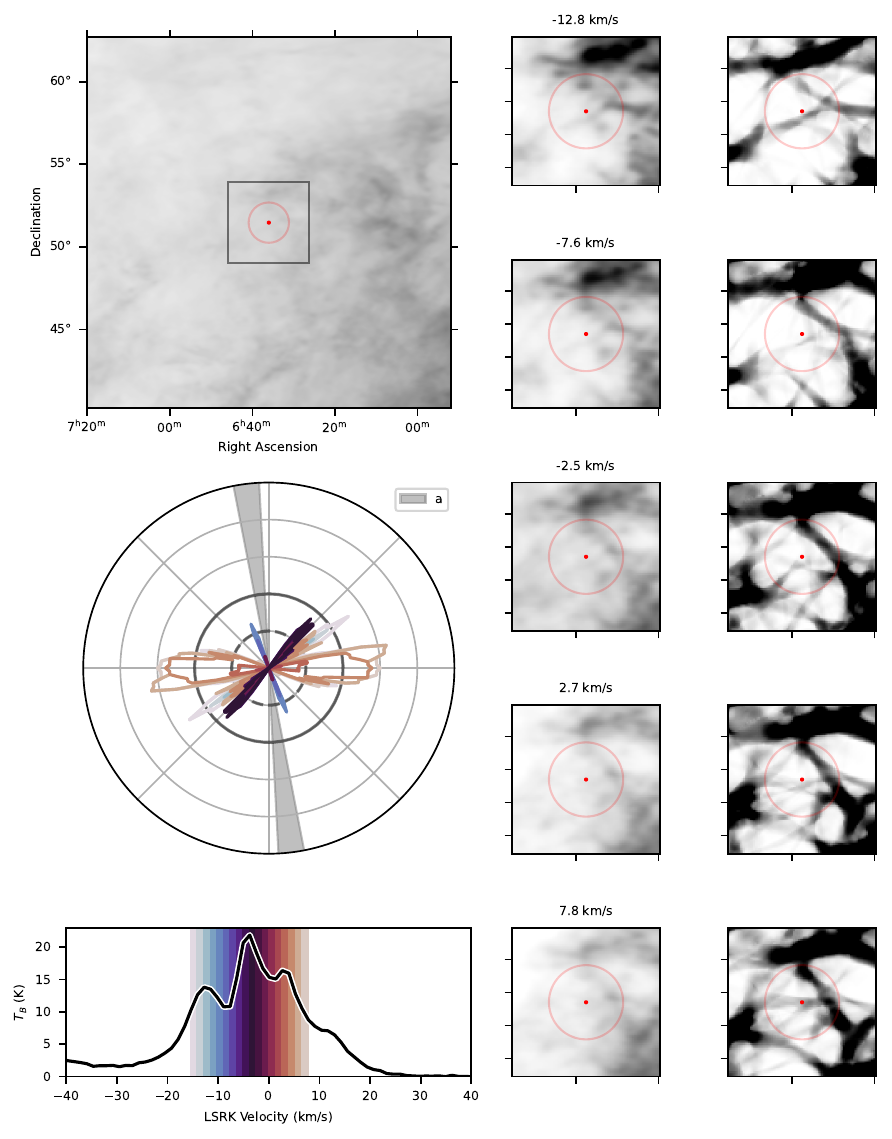}
    \caption{\HI filaments near PSR\,J0636+5128. The identified \HI filaments are not aligned with the screen.
      {\em Left column, top:\/} \HI brightness temperature integrated over velocites between $-15.4$ and $7.8{\rm\;km/s}$ relative to the kinematic local standard of rest.
      The location of the pulsar is marked with a red dot, and the window for the Rolling Hough Transform analysis with a red open circle.
      {\em Left column, middle:\/} Rolling Hough Transform spectra at the position of the pulsar for a range of velocity channels, using colours from the panel below.
      The orientation of the pulsar's scattering screen is marked with a grey shaded region.
      {\em Left column, bottom:\/} \HI line profile towards the pulsar, with a colour bar overlaid to show the velocity channels used above.
      {\em Middle column:\/} \HI brightness temperature for different velocity channels (as labelled).  The region covered is indicated by the inset box in the top left panel.
      {\em Right column:\/} Corresponding Rolling Hough Transform backprojections.\\}
\end{figure}

\begin{figure}
    \centering
    \label{fig:B0834}
    \includegraphics[width=0.8\textwidth]{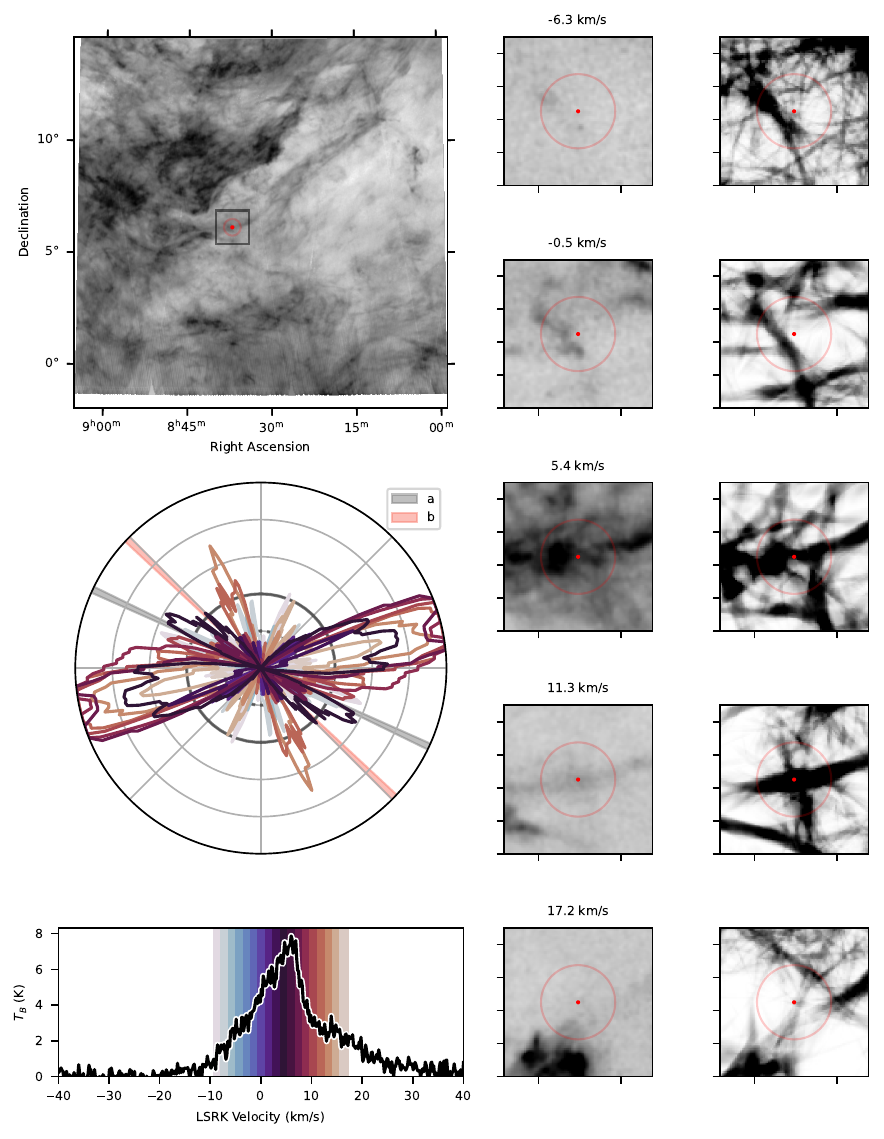}
    \caption{\HI filaments near PSR\,B0834+06. The identified \HI filaments are not aligned with either screen.
      {\em Left column, top:\/} \HI brightness temperature integrated over velocites between $-9.3$ and $17.2{\rm\;km/s}$ relative to the kinematic local standard of rest.
      The location of the pulsar is marked with a red dot, and the window for the Rolling Hough Transform analysis with a red open circle.
      {\em Left column, middle:\/} Rolling Hough Transform spectra at the position of the pulsar for a range of velocity channels, using colours from the panel below.
      The orientation of the pulsar's scattering screen is marked with a grey shaded region.
      {\em Left column, bottom:\/} \HI line profile towards the pulsar, with a colour bar overlaid to show the velocity channels used above.
      {\em Middle column:\/} \HI brightness temperature for different velocity channels (as labelled).  The region covered is indicated by the inset box in the top left panel.
      {\em Right column:\/} Corresponding Rolling Hough Transform backprojections.\\}
\end{figure}

\begin{figure}
    \centering
    \label{fig:B1133}
    \includegraphics[width=0.8\textwidth]{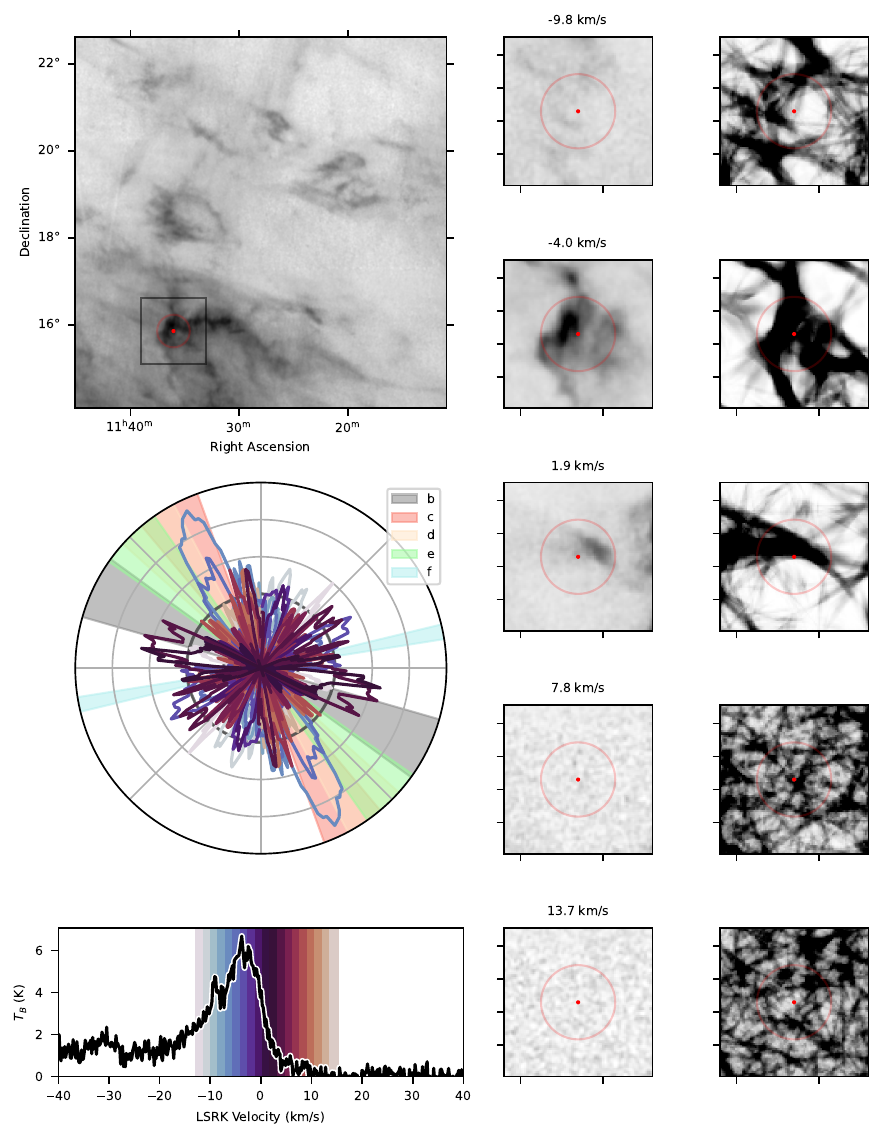}
    \caption{\HI filaments near PSR\,B1133+16. Screens c, d, and e have probable alignment with \HI filaments.
      {\em Left column, top:\/} \HI brightness temperature integrated over velocites between $-12.8$ and $15.2{\rm\;km/s}$ relative to the kinematic local standard of rest. At velocities above $7.8{\rm\;km/s}$ the emission is dominated by noise.
      The location of the pulsar is marked with a red dot, and the window for the Rolling Hough Transform analysis with a red open circle.
      {\em Left column, middle:\/} Rolling Hough Transform spectra at the position of the pulsar for a range of velocity channels, using colours from the panel below.
      The orientation of the pulsar's scattering screen is marked with a grey shaded region.      The errors in the orientations of screens c, d, e, and f are shown as $10^\circ$ which is an underestimate of the measurement uncertainties but allows for them to be more easily distinguished from each other.
      {\em Left column, bottom:\/} \HI line profile towards the pulsar, with a colour bar overlaid to show the velocity channels used above.
      {\em Middle column:\/} \HI brightness temperature for different velocity channels (as labelled).  The region covered is indicated by the inset box in the top left panel.
      {\em Right column:\/} Corresponding Rolling Hough Transform backprojections.\\
    }
\end{figure}

\begin{figure}
    \centering
    \label{fig:B1508}
    \includegraphics[width=0.8\textwidth]{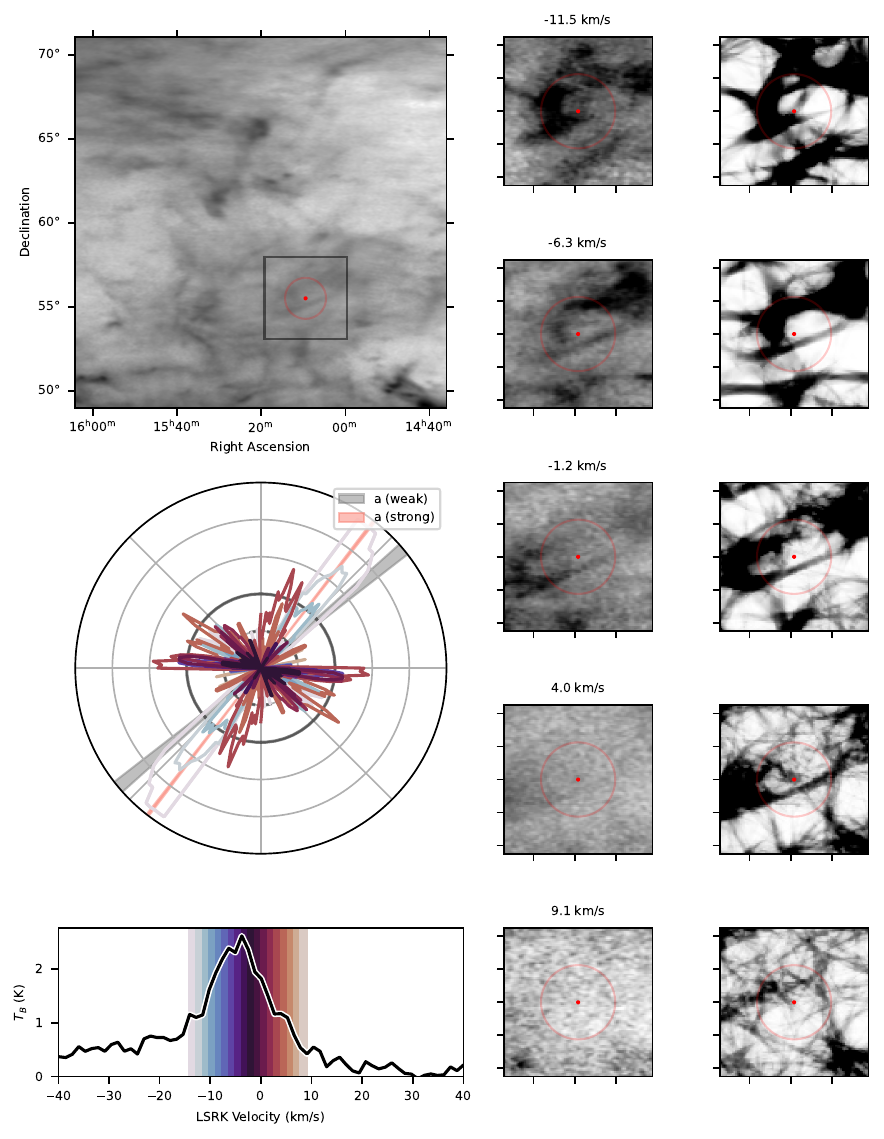}
    \caption{\HI filaments near PSR\,B1508+55. There is probable alignment of an \HI filament with strong scattering screen a, but not with the other screens.
      {\em Left column, top:\/} \HI brightness temperature integrated over velocites between $-14.1$ and $9.1{\rm\;km/s}$ relative to the kinematic local standard of rest.
      The location of the pulsar is marked with a red dot, and the window for the Rolling Hough Transform analysis with a red open circle.
      {\em Left column, middle:\/} Rolling Hough Transform spectra at the position of the pulsar for a range of velocity channels, using colours from the panel below.
      The error in the orientation of the strong scattering screen a is shown as $10^\circ$ which is an underestimate of the measurement uncertainty but allows for it to be more easily distinguished from weak scattering screen a.
      {\em Left column, bottom:\/} \HI line profile towards the pulsar, with a colour bar overlaid to show the velocity channels used above.
      {\em Middle column:\/} \HI brightness temperature for different velocity channels (as labelled).  The region covered is indicated by the inset box in the top left panel.
      {\em Right column:\/} Corresponding Rolling Hough Transform backprojections.\\
    }
\end{figure}

\begin{figure}
    \centering
    \label{fig:J1643}
    \includegraphics[width=0.8\textwidth]{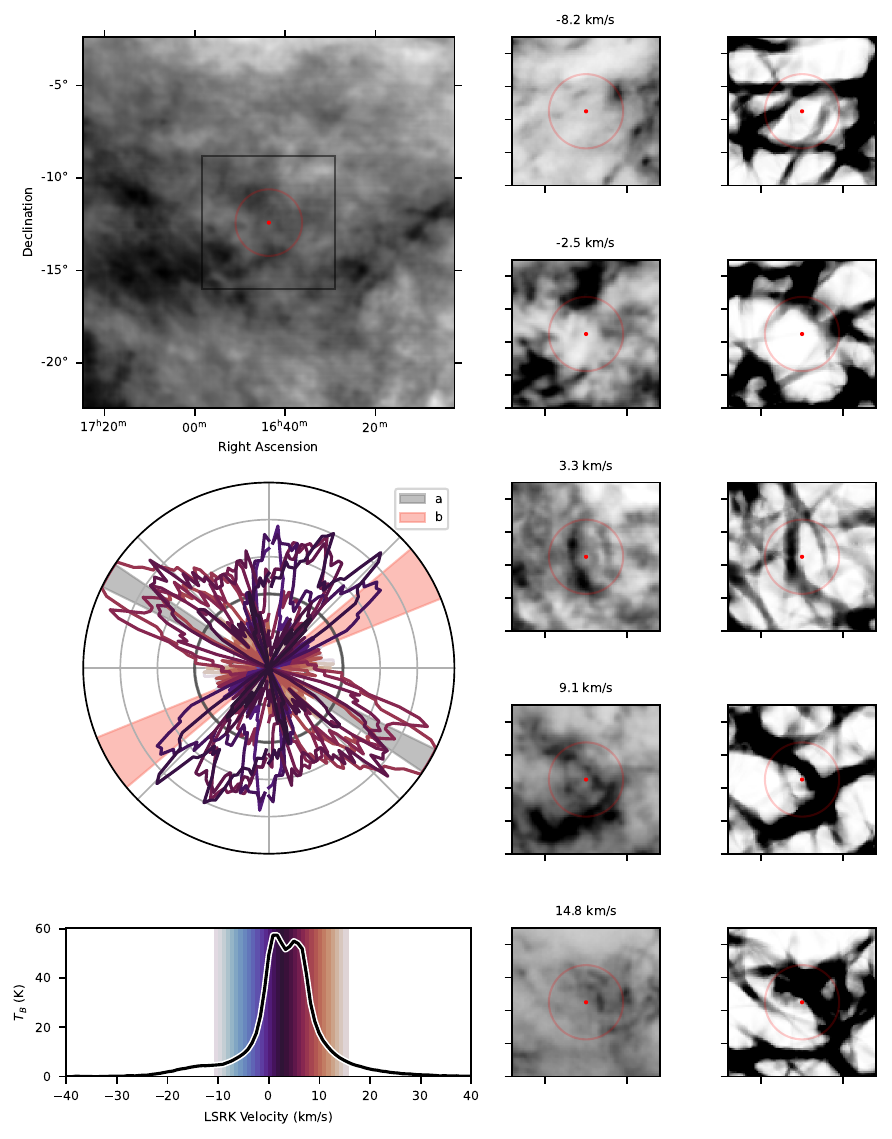}
    \caption{\HI filaments near PSR\,J1643$-$1224. Both screens have probable alignment with an \HI filament.
      {\em Left column, top:\/} \HI brightness temperature integrated over velocites between $-10.7$ and $15.7{\rm\;km/s}$ relative to the kinematic local standard of rest.
      The location of the pulsar is marked with a red dot, and the window for the Rolling Hough Transform analysis with a red open circle.
      {\em Left column, middle:\/} Rolling Hough Transform spectra at the position of the pulsar for a range of velocity channels, using colours from the panel below.
      The orientation of the pulsar's scattering screen is marked with a grey shaded region.
      {\em Left column, bottom:\/} \HI line profile towards the pulsar, with a colour bar overlaid to show the velocity channels used above.
      {\em Middle column:\/} \HI brightness temperature for different velocity channels (as labelled).  The region covered is indicated by the inset box in the top left panel.
      {\em Right column:\/} Corresponding Rolling Hough Transform backprojections.\\
    }
\end{figure}

\begin{figure}
    \centering
    \label{fig:J1909}
    \includegraphics[width=0.8\textwidth]{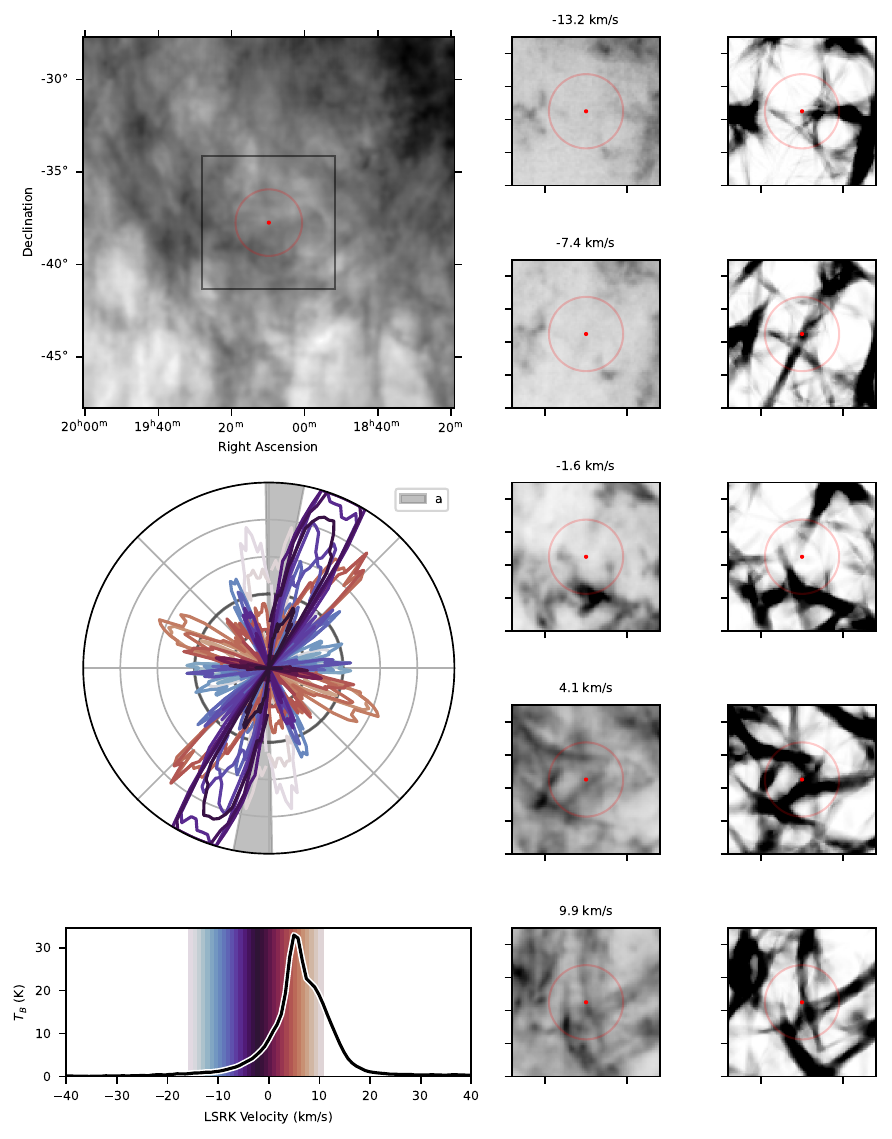}
    \caption{\HI filaments near PSR\,J1909$-$3744. The identified \HI filaments are not aligned with the screen.
      {\em Left column, top:\/} \HI brightness temperature integrated over velocites between $-15.7$ and $10.7{\rm\;km/s}$ relative to the kinematic local standard of rest.
      The location of the pulsar is marked with a red dot, and the window for the Rolling Hough Transform analysis with a red open circle.
      {\em Left column, middle:\/} Rolling Hough Transform spectra at the position of the pulsar for a range of velocity channels, using colours from the panel below.
      The orientation of the pulsar's scattering screen is marked with a grey shaded region.
      {\em Left column, bottom:\/} \HI line profile towards the pulsar, with a colour bar overlaid to show the velocity channels used above.
      {\em Middle column:\/} \HI brightness temperature for different velocity channels (as labelled).  The region covered is indicated by the inset box in the top left panel.
      {\em Right column:\/} Corresponding Rolling Hough Transform backprojections.\\
    }
\end{figure}

\end{document}